\documentclass[twocolumn,secnumarabic,amssymb,nobibnotes,aps,prl,superscriptaddress,nofootinbib]{revtex4-2}

\usepackage[utf8]{inputenc}
\usepackage{graphicx,amssymb,amsmath,amsthm,amsfonts,epsfig,epsf}
\usepackage[caption=false]{subfig}
\usepackage[usenames,dvipsnames,svgnames,table]{xcolor}
\usepackage{epstopdf}
\definecolor{darkred}{rgb}{0.5,0,0}
\usepackage{bm}
\usepackage{dcolumn}
\usepackage{latexsym}
\usepackage{rotating}
\usepackage{longtable}

\usepackage{enumerate}
\usepackage{tensor,multirow}
\usepackage{url}
\usepackage[linktocpage]{hyperref}

\hypersetup{
	colorlinks,
	linkcolor={red!60!black},
	citecolor={green!50!black},
	urlcolor={blue!80!black}
}

\def\nn{\nonumber}

\def\be{\begin{equation}}
\def\ee{\end{equation}}
\newcommand{\beq}{\begin{eqnarray}}
\newcommand{\eeq}{\end{eqnarray}}
\def\ba{\begin{align}}
\def\ea{\end{align}}

\begin{document}
	
\title{Parasitic black holes: the swallowing of a fuzzy dark matter soliton}

\author{Vitor Cardoso}
\affiliation{CENTRA, Departamento de F\'{\i}sica, Instituto Superior T\'ecnico -- IST, Universidade de Lisboa -- UL,
	Avenida Rovisco Pais 1, 1049 Lisboa, Portugal}
\affiliation{Niels Bohr International Academy, Niels Bohr Institute, Blegdamsvej 17, 2100 Copenhagen, Denmark}
\author{Taishi Ikeda} 
\affiliation{Dipartimento di Fisica, ``Sapienza'' Universit\'{a} di Roma, Piazzale Aldo Moro 5, 00185, Roma, Italy}
\author{Rodrigo Vicente}
\affiliation{Institut de Fisica d’Altes Energies (IFAE), The Barcelona Institute of Science and Technology, Campus UAB, 08193 Bellaterra (Barcelona), Spain}
\author{Miguel Zilhão}
\affiliation{Departamento de Matem\'atica da Universidade de Aveiro and
	Centre for Research and Development in Mathematics and Applications (CIDMA),
	Campus de Santiago, 3810-183 Aveiro, Portugal}
%

\begin{abstract}
Fuzzy dark matter is an exciting alternative to the standard cold dark matter paradigm, reproducing its large scale predictions, while solving most of the existing tension with small scale observations. These models postulate that dark matter is constituted by light bosons and predict the condensation of a solitonic core -- also known as boson star, supported by wave pressure -- at the center of halos.
However, solitons which host a \emph{parasitic} supermassive black hole are doomed to be swallowed by their guest. It is thus crucial to understand in detail the accretion process. In this work, we use numerical relativity to self-consistently solve the problem of accretion of a boson star by a central black hole, in spherical symmetry. We identify three stages in the process, a {\it boson-quake}, a {\it catastrophic stage} and a linear phase, as well as a general accurate expression for the lifetime of a boson star with an endoparasitic black hole. Lifetimes of these objects can be large enough to allow them to survive until the present time.
\end{abstract}

\maketitle

\noindent {\textbf{Introduction.}}\label{sec:intro}
One of the most solid predictions of the fuzzy dark matter model (FDM) is that coherent solitonic cores condense at the center of virialized FDM halos, satisfying the soliton-halo mass relation~\cite{Schive:2014dra,Schive:2014hza,Veltmaat:2018dfz}
\begin{equation} 
M_\text{BS}\approx 6.5 \times 10^{9} M_\odot\, m_{22}^{-1} \left(\frac{M_\text{halo}}{10^{14} M_\odot}\right)^{\frac{1}{3}}\,,\label{core-halo}
\end{equation}
while the outer halo profile resembles the Navarro-Frenk-White profile for cold dark matter (CDM) halos~\cite{Navarro:1995iw}. Here, $m_{22}\equiv m_\psi/10^{-22}\, \text{eV}$, where~$m_\psi$ is the boson mass. These solitons are self-gravitating configurations of a scalar field supported by wave pressure, described well by ground-state stationary boson stars (BSs)~\cite{Kaup:1968zz,Ruffini:1969qy,Friedberg:1986tp,Lee:1991ax,Liebling:2012fv,Visinelli:2021uve} (for complex scalars), or long-lived oscillatons~\cite{Bogolyubsky:1976nx,Seidel:1991zh,Copeland:1995fq,Page:2003rd,Urena-Lopez:2001zjo} (for real scalars). They can form through \emph{gravitational cooling}~\cite{Seidel:1993zk,Guzman:2006yc}; it was argued that this mechanism may be understood in terms of two-body relaxation of wave granules over a timescale~\cite{Hui:2016ltb,Hui:2021tkt} (see also Refs.~\cite{Levkov:2018kau,Bar-Or:2018pxz,Bar-Or:2020tys}) 
\begin{equation}
t_\text{relax}\sim 10^8 \, \text{yr} \bigg(\frac{R}{2 \, \text{kpc}}\bigg)^4 \Big(\frac{v}{100\,\text{km/s}}\Big)^2 m_{22}^3\,,
\end{equation}
for a typical galactic DM velocity~$v$ and for a relaxed region of radius~$R\sim 2 \,\text{kpc}$. Assuming that the relation~\eqref{core-halo} holds, for given host halo of mass~$M_\text{halo}$, the density profile of a FDM soliton is entirely determined by the boson mass~$m_\psi$. Using galactic rotation curves from the SPARC database~\cite{Lelli:2016zqa}, stringent constraints on~$m_\psi$ can be imposed~\cite{Bar:2018acw,Bar:2019bqz,Bar:2021kti}. In particular, these results disfavor FDM with~$10^{-24} \, \text{eV} \lesssim m_\psi \lesssim 10^{-20}\, \text{eV}$ from comprising all DM; similar type of constraints were found from the stellar orbits near Sgr\,A* and by combining stellar velocity measurements with the Event Horizon Telescope imaging of M87*~\cite{Bar:2019pnz}.
Most of these studies are based on the assumption that the soliton mass and profile remains largely unaltered since its formation.\footnote{There are also important cosmological constraints from, e.g., the Lyman-$\alpha$ forest~\cite{Irsic:2017yje,Kobayashi:2017jcf,Armengaud:2017nkf,Zhang:2017chj,Rogers:2020ltq} and the cosmic microwave background anisotropy~\cite{Hlozek:2017zzf} which do not resort to this assumption.}

However, there is strong evidence that all large galaxies (like our own Milky Way) or even dwarf galaxies possess a central supermassive black hole (SMBH)~\cite{Kormendy:1995,Kormendy:2013dxa,Reines:2022}. So, FDM solitons are expected to host a \emph{parasite} SMBH feeding from it, growing and, eventually, swallowing it, as suggested by no-hair results~\cite{Ruffini:1971bza,Israel:1967wq,Israel:1967za,Carter:1971zc,Bekenstein:1972ny}.
Despite this, most studies in the literature \emph{neglect} the effect of SMBHs on solitons. 
The rationale for doing so is often based on approximation schemes to estimate the impact of BH accretion on the soliton, either by using the BH absorption cross-section~\cite{Hui:2016ltb,Bar:2018acw,Bar:2019pnz} (formally only well-defined for scattering states, whereas BSs are bounded), by using the decay rate of ``gravitational atom'' states (valid only when the BH dominates the dynamics)~\cite{Urena-Lopez:2002nup,Barranco:2011eyw,Barranco:2012qs,Barranco:2013rua,Davies:2019wgi,Brax:2020tuk}, or by evolving numerically the system, but for short timescales and with fine-tuned initial data~\cite{Barranco:2017aes}. BH accretion of diffuse scalars was also studied in~\cite{Cruz-Osorio:2010nua,Urena-Lopez:2011lcm,Guzman:2012jc,Hui:2019aqm,Clough:2019jpm,Bamber:2020bpu}. While the different schemes predict quite disparate scalings for the accretion timescale, all of them suggest that for typical FDM masses this timescale is larger than a Hubble time.
None of the existing treatments in the literature captures the full picture of BS accretion by SMBHs. 

Here, we use numerical relativity to evolve the full system  -- in spherical symmetry -- for long timescales, covering the whole accretion process, and find general accurate expressions for the accretion time.  We adopt the mostly positive metric signature and use geometrized units ($c=G=1$).

\noindent{\textbf{Setup.}} 
Consider a complex scalar field~$\psi$ minimally coupled to the spacetime metric~$g_{\mu\nu}$ described by the action
\begin{equation}
	S=\int d^4x \sqrt{-g}\left(\frac{R}{16 \pi}-\nabla_\mu \psi \nabla^\mu \psi^*-\mu^2 |\psi|^2\right),
\end{equation}
where~$R$ is the scalar curvature,~$g\equiv \det\left(g_{\mu\nu}\right)$ is the metric determinant, and~$\mu\equiv m_\psi/\hbar$ is the inverse of the reduced Compton wavelength. The first variations of the action yield the Einstein-Klein-Gordon field equations
\beq
R_{\mu \nu} -\frac{1}{2}g_{\mu \nu}R=8\pi T_{\mu \nu}\,,\qquad (\Box-\mu^2) \psi=0,\label{EKG}
\eeq	
where~$R_{\mu \nu}$ is the Ricci tensor and~$\Box \equiv g^{\mu \nu} \nabla_\mu \nabla_\nu$ is the covariant d'Alembert operator, with the energy-momentum tensor $
T_{\mu\nu}=\partial_{\mu}\psi\partial_{\nu}\psi^{*}+\partial_{\nu}\psi\partial_{\mu}\psi^{*}-g_{\mu\nu}(
\partial_\alpha\psi\partial^\alpha \psi^{*}+\mu^{2}\left|\psi\right|^{2})$.
We shall describe the scalar particles through the classical field~$\psi$, since the average particle number $N_\text{BS}$ in a FDM soliton is extremely large,
\begin{equation}
N_\text{BS}\sim 10^{97}\,m_{22}^{-1}\frac{M_\text{BS}}{10^{9} M_\odot}\,,
\end{equation}
and quantum fluctuations (in a coherent state) are negligible for a large average occupation number~\cite{Glauber:1963fi,Hui:2021tkt}.

The FDM soliton will be described by a ground-state spherically symmetric BS~\cite{Li:2020ryg}, which are regular stationary solutions of equations \eqref{EKG} with~$\psi =\chi(r) e^{-i \Omega t}$. 
For a mass $M_\text{BS}\lesssim  4 \times 10^{11}M_\odot\, m_{22}^{-1}$
(equivalently, central amplitude~$\chi(0)\lesssim 10^{-2}$), they are Newtonian objects and described through the simpler Poisson-Schrödinger system, the Newtonian limit of system \eqref{EKG}~\cite{Annulli:2020lyc}. In this limit, ground-state BSs satisfy the mass-radius relation
\begin{equation}
M_\text{BS}	\approx 9 \times  10^{8}M_\odot \frac{1\, \text{kpc}}{R_\text{98}}\,m_{22}^{-2},
\end{equation}
where~$R_{98}$ is the radius enclosing~$98\%$ of~$M_\text{BS}$,
and oscillate with frequency~$\sim \mu/2 \pi\approx 0.76\, \text{yr}^{-1} \, m_{22}$.
These objects are stable under linear perturbations and their fundamental normal mode frequency is~\cite{Annulli:2020lyc,Guzman:2004wj}
\begin{equation} \label{nm}
\frac{\omega_\text{NM}}{2\pi}\approx 0.029\,\text{Myr}^{-1}\,m_{22}^2\left(\frac{M_\text{BS}}{10^{9} M_\odot}\right)^2\,.
\end{equation}
For simplicity, we consider a spherically symmetric system at all times. We focus on initial data describing BHs with mass~$M_\text{BH,0}\equiv M_\text{BH}(t=0)$ smaller than the BS mass~$M_\text{BS,0}$. The full system~\eqref{EKG} is evolved using numerical relativity (our numerical scheme and initial data are described in the Supplemental Material).

\noindent{\textbf{Adiabatic approximation.}} 
We consider first a Newtonian BS and use an adiabatic approximation (see also Refs.~\cite{Urena-Lopez:2002nup,Barranco:2017aes}), which is useful to understand our numerical results.
Assume that the BS mass changes at a much smaller rate than~$\mu/2\pi$, so that the field is~$\psi\approx\chi(r) e^{-i \left(\mu-i \gamma(t)\right) t}$ with~$0<|\gamma|\ll \mu$, for~$|t|\ll \min \big(|\gamma/\partial_t \gamma|,|\gamma/\partial_t^{2} \gamma|^{\frac{1}{2}}\big)$. 
Within the BH influence radius~$r_i=M_\text{BH}/ |U_\text{BS}(r_i)|\sim \tfrac{1}{2} \left(M_\text{BH}/M_\text{BS}\right) R_{98}$ 
(where~$U_\text{BS}$ is the BS gravitational potential) one can use the test field approximation, describing the field through the Klein-Gordon equation on a Schwarzschild background; the radial field is then~\cite{Starobinski:1973,Vicente:2022ivh}
\begin{equation}\label{near_sol}
\chi \approx \begin{cases}
A\,e^{-2i\mu M_\text{BH} \log\left(1-\frac{2M_\text{BH}}{r}\right)},  \quad r \ll 1/\mu\\
\frac{A \xi}{i C_0 r} \left( \text{F}_0-\frac{4C_0^2\mu M_\text{BH}^2}{\xi}\, \text{G}_0\right), \;\;  2M_\text{BH}\ll r \lesssim r_i \nonumber
\end{cases}
\end{equation}
where~$\text{F}_0(\eta,i\frac{r}{\xi})$ and $\text{G}_0(\eta,i\frac{r}{\xi})$ are Coulomb functions~\cite{NIST:DLMF}. We define~$C_0\equiv \left|\Gamma\left(1+i \eta\right)\right| e^{-\eta \pi/2}$ and~$\eta\equiv i \mu^2 M_\text{BH} \xi$, with~$\xi\equiv 1/\sqrt{2 i \mu \gamma}$ and~$\text{Re}\, \xi>0$, where~$\Gamma$ is the gamma function. For~$r\gg 2 M_\text{BH}$, the field satisfies 
\begin{subequations}\label{dirtyBS}
	\begin{gather}
	\partial_r\left(r^2\partial_r \chi\right)\approx r^2\bigg[\frac{1}{\xi^2}+2\mu^2 \bigg(U_\text{BS}-\frac{M_\text{BH}}{r}\bigg)\bigg]\chi, \\
	\partial_r\left(r^2\partial_r U_\text{BS}\right)\approx 8\pi \mu^2 r^2\left|\chi\right|^2,
	\end{gather}
\end{subequations}
describing a ``dirty'' BS distorted by the BH gravitational field.
In general, the above system is a boundary value problem with complex eigenvalue~$\xi$ that must be solved numerically. The overall scale factor~$A$ is determined by the condition that the total mass of the field is~$M_\text{BS}$. 

For~$\nu \equiv M_\text{BH}/M_\text{BS}\lesssim 1/6$, one can show that~\cite{Annulli:2020lyc}
\begin{gather} 
	A\approx 4.7\times 10^{-2}\mu^2 M_\text{BS}^2\left(1+6 \nu\right), \\
	\text{Im}\, \gamma \sim - 10^{-1}\, \mu^3 M_\text{BS}^2\,.
\end{gather}
One can use this expression for~$A$ to compute the flux of energy through the horizon and find the rate of accretion $\dot{M}_{\rm BH}\equiv d M_\text{BH}/dt$,
\begin{equation} \label{adiab_i}
\dot{M}_{\rm BH} \approx 32 \pi \left[4.7\times 10^{-2} \mu^3 M_\text{BH} M_\text{BS}^2(1+6 \nu)\right]^2\,,
\end{equation}
which can then be solved numerically, using energy conservation $M_\text{BS}=M_\text{BS,0}+M_\text{BH,0}-M_\text{BH}$. Also by energy conservation,~$2 M_ \text{BS}\text{Re}\,\gamma=\dot{M}_{\rm BH}$, or
\begin{equation}
	\frac{\text{Re}\,\gamma}{\mu} \approx 0.1 \mu^5 M_\text{BH}^5 \left(1+6 \nu\right)^2/\nu^{3}\ll 1,
\end{equation}
which is consistent with the adiabatic assumption.

For~$\nu\gtrsim 2$, one has~$r_\text{i}\approx R_\text{98}$, implying that the test field approximation is valid almost everywhere, and so the scalar field is described by a superposition of \emph{gravitational atom} states~\cite{Detweiler:1980uk,Baumann:2019eav,Ikeda:2020xvt} [$\chi\equiv \sum_n c_n \chi_ n$ with~$\sum_n |c_n|^2=1$], with
\be
A_n\approx  \frac{\mu^2  M_\text{BH}^{2}}{\sqrt{2\pi \nu}\,n^{3/2}}\,,\qquad
\text{Im}\, \gamma_ n \approx - \frac{\mu^3 M_\text{BH}^2}{2 n^2}\,,\label{An} 
\ee
with the integer~$n\geq1$. 
Although weak, the self-gravity of the scalars is responsible for~$R_{98}\lesssim \text{Re}\,\xi_1$, so that most support is expected to be in the state~$n=1$. This is consistent with the projection of a ground-state BS onto gravitational atom states (Supplemental Material).
Using the analytic expression~\eqref{near_sol} to compute the flux of energy through the event horizon gives
\begin{equation} 
\dot{M}_{\rm BH}\approx 16 \mu^6 M_\text{BH}^5 M_\text{BS}\,,\label{adiab_ii}
\end{equation}
which can be solved numerically for~$M_\text{BH}$, and implies
\begin{equation}
	\frac{\text{Re}\,\gamma_1}{\mu} \approx 8 \mu^5 M_\text{BH}^5\ll 1\,.\label{gr_atom_decay}
\end{equation}
The instantaneous decay rate is in clear agreement with Refs.~\cite{Detweiler:1980uk,Brito:2015oca,Baumann:2019eav} and is consistent with adiabaticity.

\noindent{\textbf{Numerical results.}} 
%
\begin{figure}[thbp]
	\centering
	\includegraphics[width=0.53\textwidth]{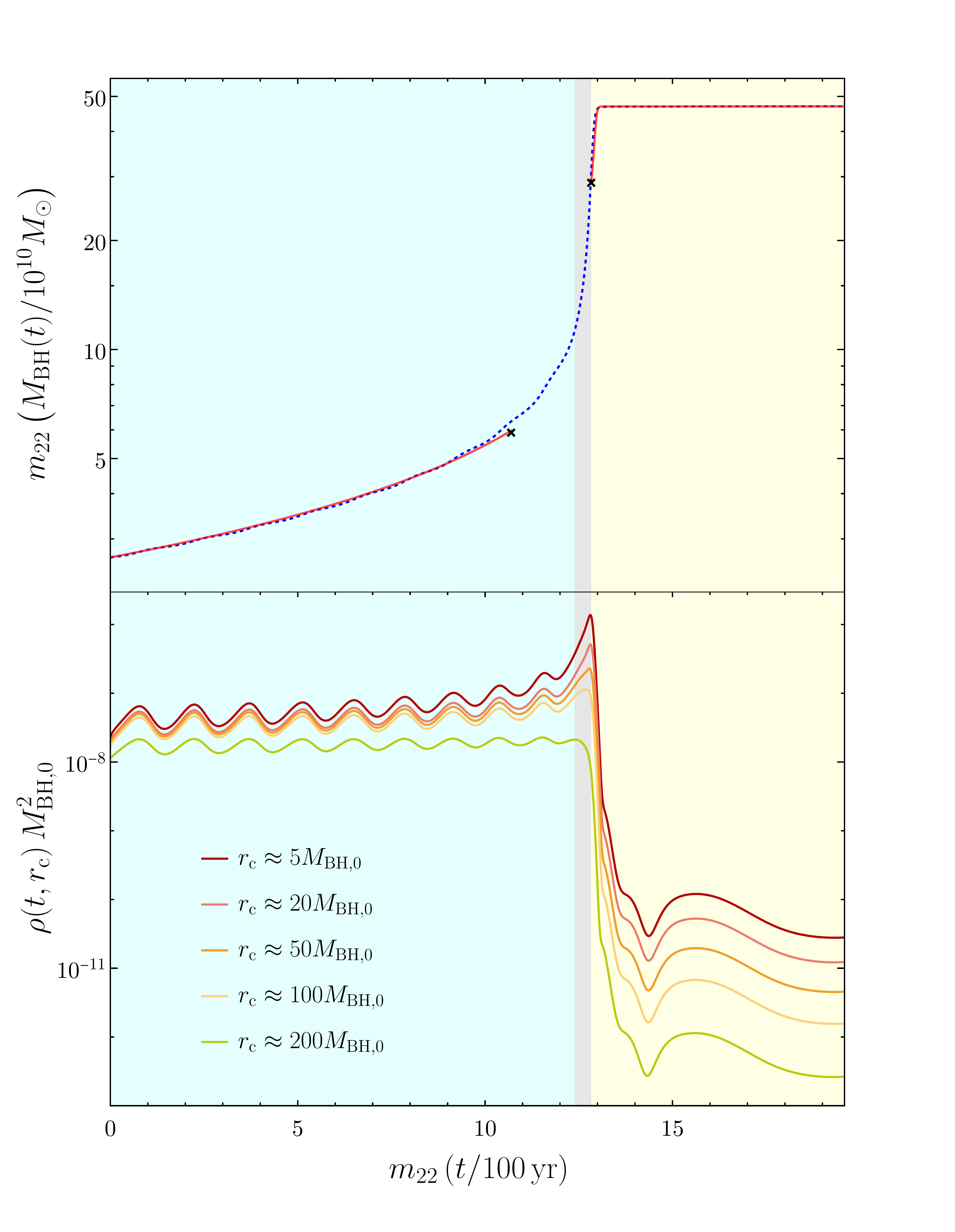}
	\caption{Stages of accretion of a FDM soliton by a endoparasitic SMBH. Blue corresponds to a boson-quake -- the  excitation of BS modes by the accreting BH, stage~I in the process. Gray corresponds to a violent accretion process, stage~II. In stage III, the BH dominates the entire dynamics, and the spacetime is well described by a slightly perturbed BH, in yellow (notice the cascade starting in phase III, which indicates a dominance of progressively higher modes). {\bf Top:} BH mass as function of time (dashed blue) for initial masses~$(M_\text{BS,0},M_\text{BH,0})\approx (40,3)\times 10^{10}M_\odot \,m_{22}^{-1}$. Red curves show analytical approximation~\eqref{adiab_i} and~\eqref{adiab_ii}, black crosses signal~$\nu=\{1/6,2\}$ (where, respectively,~\eqref{adiab_i} ceases and~\eqref{adiab_ii} starts to be valid). 
{\bf Bottom:} Energy density of scalar field as function of time at several different radii~$r_\text{c}$.
}
	\label{accr_stages}
\end{figure}
Using numerical relativity we can track the entire evolution of both the central SMBH and the soliton (the BH mass is computed from the apparent horizon area; the initial data construction and numerical scheme employed follow standard approaches~\cite{Alcubierre:2004gn,Arbona:1998hu,Shibata:1995we,Baumgarte:1998te,Gundlach:2006tw,Cardoso:2014uka,Brown:2007nt,Montero:2012yr,Yo:2002bm} and are detailed in the Supplemental Material). We studied BS-BH systems with mass ratios up to~20 and size ratio up to~$10^3$, probing the limits of our numerical scheme.
Figure~\ref{accr_stages} shows the results of one particular simulation with parameters~$M_\text{BS,0}\approx 4\times 10^{11}M_\odot \,m_{22}^{-1}$ and~$\nu_0 \approx 1/16$. In the top panel, we show the evolution of the BH mass and in the bottom panel the energy density of the scalar field measured at different radii~$r_\text{c}$.
All of our simulations are characterized by three main stages of accretion (which we label as~I, II and III) that we now describe. 
Results for different initial parameters can be found in the Supplemental Material.

In stage I the dynamics is controlled mainly by the soliton, since the scalar field amplitude close to the horizon depends strongly on the BS self-gravity. The initial data for the scalar describes a ``pure'' BS, while the quasi-equilibrium configuration is a ``dirty'' BS. Thus, when the simulation starts, a boson-quake is excited and the soliton oscillates with frequency $\sim\omega_\text{NM}$ around an equilibrium ``dirty'' BS that evolves adiabatically. These oscillations are clearly seen in the energy density of the field, and are also present in the evolution of the BH mass. The accretion rate in this stage is very well described by the analytic approximation~\eqref{adiab_i}, at least until~$\nu\approx1/6$; after that point, the analytical model tends to underestimate the accretion rate. We define the end of stage I to be the instant when~$\mu M_\text{BH}= 0.08$ or~$\nu=2$ is attained, whichever happens first.

If the BH becomes massive enough that~$\mu M_\text{BH}\geq 0.08$, but still with mass ratio~$\nu<2$, the system enters stage~II. This ``catastrophic'' stage of accretion is
triggered by a very efficient tunneling of the field through the potential barrier~\cite{Bamber:2020bpu} (the maximum in the effective potential disappears at~$\mu M_\text{BH}= 0.25$). This stage lasts for one free-fall time~$\tau_\text{FF} \sim [R_\text{98}^3/(M_\text{BH}+M_\text{BS})]^\frac{1}{2} $, during which the BH mass grows exponentially.
We define the end of stage II to occur when $\nu=2$.

When the BH grows to~$\nu\gtrsim 2$, the whole dynamics is controlled by the BH. In this stage, the BH influence radius is of the order of the configuration size, which implies that the whole scalar behaves as a test field on a Schwarzschild spacetime, whose mass evolves adiabatically. This picture is confirmed by the fact that the accretion rate is very well described by the analytic expression~\eqref{adiab_ii}. The BH mass saturates at~$\sim M_\text{BS,0}+M_\text{BH,0}$, which is compatible with none of the scalar being radiated away. At late times, the field decays in a superposition of states, starting at the $n=1$ which is the {\it shortest-lived} mode, cf. Eq.~\eqref{An}. Thus a ``peeling-off'' of different modes is apparent in Fig.~\ref{accr_stages}, which was also seen recently during the collision between BHs and BSs~\cite{Cardoso:2022vpj}. At very late times, a power-law decay will settle in~\cite{Koyama:2001ee,Witek:2012tr}, but a clear imprint would require prohibitively large timescales.

\begin{figure}[thbp]
	\centering
	\includegraphics[width=0.45\textwidth]{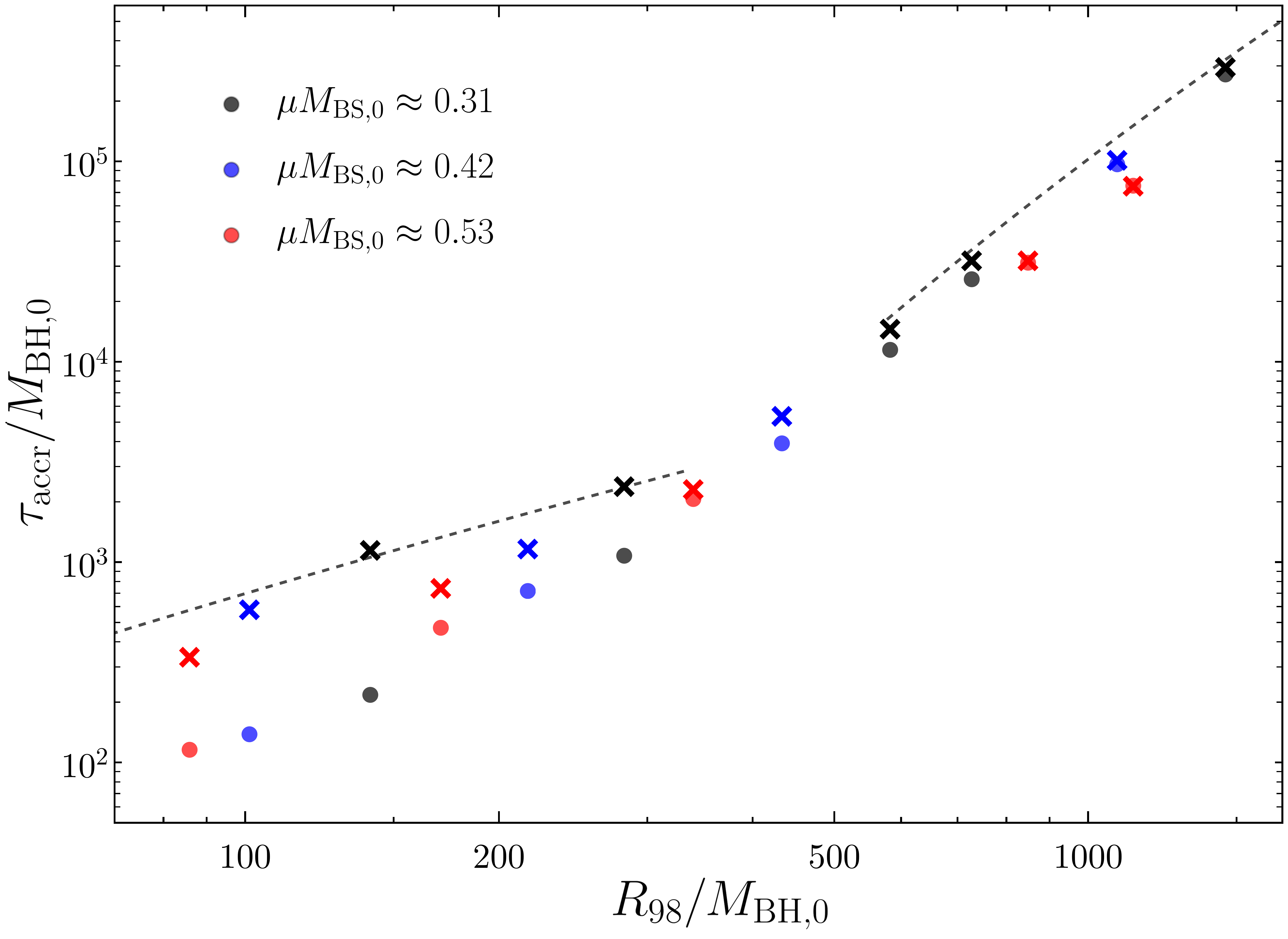}
	\caption{BS accretion time as function of~$R_{98}/M_\text{BH,0}$ for different configurations. Dots (crosses) represent~$\tau_{10\%}$ ($\tau_{90\%}$), the time for~$10 \%$ ($90\%$) of the BS mass to be accreted by the BH. For given BS mass (fixed color), points to the left have larger~$\nu_0\sim \mathcal{O}(1)$, while points to the right have smaller~$\nu_0\lesssim 1/6$. Black dashed lines show the analytical prediction for $\mu M_\text{BS,0}\approx 0.31$: left is free-fall time~$\tau_\text{FF}\approx [R_{98}^3/(M_\text{BH,0}+M_\text{BS,0})]^\frac{1}{2}$, and right is Eq.~\eqref{accr_time_BS} for~$\tau_{10\%}$. Even though this BS is only marginally Newtonian, the agreement is remarkable (note: we are not fitting any free parameter).
	}
	\label{accr_time}
\end{figure}

Figure~\ref{accr_time} shows the accretion timescales $\tau_{10\%}$ (dots), $\tau_{90\%}$ (crosses) for different initial configurations, defined as the time for~$10\%,\,90\%$ of the soliton mass to be accreted by the BH, respectively. As discussed above, in all our simulations most of the soliton mass is accreted during stage II, which lasts a free-fall time~$\tau_\text{FF}$; thus, the difference between~$\tau_{10\%}$ and~$\tau_{90\%}$ is usually of the same order of~$\tau_\text{FF}$. Points to the left represent configurations with larger~$\nu_0\equiv \nu (t=0)=\mathcal{O}(1)$, implying that their accretion process do not have stage I (or else it is very short), starting already at stage II. This explains why~$\tau_{90\%}$ in the left is very well described by~$\tau_\text{FF}$, and why the relative difference between~$\tau_{10\%}$ and~$\tau_{90\%}$ is larger in this region of the plot. The points to the right represent configurations with smaller~$\nu_0$ ($\lesssim 1/6$), which have a long stage~I (longer than~$\tau_\text{FF}$). This explains why the relative difference between~$\tau_{10\%}$ and~$\tau_{90\%}$ is smaller and why~$\tau_{10\%}$ is very well described by the analytical expression~$\eqref{accr_time_BS}$ in this region of the plot. The agreement between the numerical results and the analytical expressions is remarkable.
%

\noindent{\textbf{Discussion.}} 
The accretion of a self-gravitating scalar structure by a BH in a spherically symmetric setting is perhaps the simplest dynamical process that
one can conceive of. This is the counterpart of Bondi accretion~\cite{Bondi:1952ni} for fundamental fields, hence a process which is clearly interesting from the 
physical point of view. Although we focus this discussion mainly on FDM, our results are general and have a much broader range of applications in theoretical physics.

Our simulations show that if initially a host BS is much heavier than a newborn BH ($\nu_0\lesssim 1/6$), the process starts in stage I, a slow accretion stage where the soliton dominates the dynamics and its normal modes are excited. The same type of normal mode excitation was seen in cosmological evolutions of halos~\cite{Veltmaat:2018dfz}. 
The excitation amplitude depends on the initial data and, in particular, how the BH forms (a detailed modeling of which is out of the scope of this work). 
Our results suggest that, for initial configurations with~$\nu_0\lesssim 1/6$, the accretion time is of the same order of the duration of stage I itself, which can be estimated by 
\begin{gather} 
	\frac{\tau_\text{10\%}}{10 \,\text{Gyr}} \approx 3 f(\nu_ 0) \left(\frac{10^{10} M_\odot}{M_\text{BS,0}}\right)^{5}\,m_{22}^{-6}\, , \label{accr_time_BS}\\
	f\approx \tfrac{60-470 \nu_ 0}{47 \nu_0}+\tfrac{50 \nu_0 \left(573+5530 \nu_0\right)}{282\left[1+4 \nu_0 \left(5+31 \nu_0\right) \right]}-10\log \left(\tfrac{1+22\nu_0}{15 \nu_0(1+4 \nu_ 0)}\right)\nn,	
\end{gather}
where this expression is obtained by integrating~\eqref{adiab_i}, and~$f$ is a strictly decreasing function with~$f(1/6)\approx2.2$. Note that, for these configurations,~$\tau_{90\%}\sim \tau_{10\%}$.

On the other hand, if the initial BH mass is comparable to (or larger than) the BS mass ($\nu_0\gtrsim \mathcal{O}(1)$), the process starts in stage II, a ``catastrophic'' stage where most of the BS is accreted in one free-fall time (see Supplemental Material); in this case, our results indicate that~$\tau_{90\%}$ is well described by the free-fall time (cf. Fig.~\ref{accr_time})
\begin{equation} \label{ff_t}
	\frac{\tau_\text{FF}}{10\, \text{Gyr}}\approx 10^2 \frac{(\kappa/10)^\frac{3}{2}}{(1+\nu_0^{-1})^2}\left(\frac{10^{8}M_\odot}{M_\text{BH,0}}\right)^2\,m_{22}^{-3},
\end{equation}
where~$\kappa\equiv \mu^2 R_\text{98} (M_\text{BH,0}+M_\text{BS,0})$ satisfies~$3.8\lesssim\kappa \lesssim 9.1$.
However, stage II may not exist if the initial configuration is not sufficiently massive, i.e., if~$\mu (M_\text{BH,0}+M_\text{BS,0})\ll 0.08$ (equivalently,~$m_{22}(M_\text{BH,0}+M_\text{BS,0})\ll 10^{11}M_\odot$), in which case the BH effective potential is strong enough to suppress accretion~\cite{Bamber:2020bpu}. In those cases, the distinction between different stages is highly blurred, and the process may be well described by stage III only; we have not probed this regime as it requires prohibitively large resources. If true, this picture suggests that for light configurations
with~$\nu_0\gtrsim \mathcal{O}(1)$, the accretion process is entirely linear, corresponding to gravitational atom states~\cite{Barranco:2011eyw,Barranco:2012qs,Barranco:2013rua,Davies:2019wgi} and which decay exponentially on a timescale~$\sim 5 \times 10^{18} \,\text{yr}\,(10^8 M_\odot/M_\text{BH,0})^5 m_{22}^{-6}$ (cf. Eq.~\eqref{gr_atom_decay}).

Our numerical results and analytical expressions for the accretion time of a BS hosting a parasitic BH
establish once and for all the details of the accretion of light scalars onto BHs. 
We find remarkable agreement between analytical estimates and full numerical relativity simulations for different initial configurations.
The lightest soliton we evolved has a mass~$M_\text{BS,0}\approx 4\times 10^{11}M_\odot\, m_{22}^{-1}$, considerably heavy for FDM cosmology~\cite{Kulkarni:2020pnb}. The extrapolation of our results to lighter solitons is well grounded, since our analytical expressions were derived in the Newtonian limit, and are expected to be more accurate for lighter configurations. Although we neglected the effect of spin, it is easy to show that our adiabatic approximation can be extended to a spinning BH; for $M_\text{BH}\ll 10^{11} M_\odot \,m_{22}^{-1}$, spin suppresses the accretion rate by a factor~$\left(1+\sqrt{1-(J_\text{BH}/M_\text{BH}^2)^2}\right)/2$, where~$J_\text{BH}$ is the BH angular momentum~\cite{Vicente:2022ivh}. However, for complex scalars, new ``hairy'' BH solutions exist and could be a possible endstate of the accretion process~\cite{Herdeiro:2014goa,Herdeiro:2015waa}; further study is required to understand the system away from spherical symmetry.

Taken together with relation~\eqref{core-halo}, our main result Eq.~\eqref{accr_time_BS} (note that Eq.~\eqref{ff_t} applies only to very massive BHs) implies that the lifetime of FDM cores hosting a central BH born with mass~$M_\text{BH,0}\lesssim 10^6 M_\odot$ in a halo with~$M_\text{halo}\lesssim 10^{15}M_\odot$ is larger than a Hubble time for $m_\psi\lesssim8 \times 10^{-20}\, \text{eV}$.
Thus, for an interesting region of the parameter space, FDM solitons can survive until the present day and help solve the potential small scale problems of CDM~\cite{Weinberg:2013aya}. 
However, this conclusion relies heavily on the soliton-halo relation, which neglects baryonic effects and was tested numerically only for~$M_\text{halo}\sim(10^{8},10^{11})\,M_\odot\, m_{22}^{-1}$. The strong dependence of Eq.~\eqref{accr_time_BS} on~$M_\text{BS,0}$ implies that, if the presence of baryons increases the soliton mass by a factor of two relative to~\eqref{core-halo} (as found for stars~\cite{Chan2018}), the soliton can only survive one Hubble time if~$m_\psi\lesssim 2 \times 10^{-22}\, \text{eV}$.

\noindent {\bf \em Acknowledgments.} 
We are grateful to Fabrizio Corelli for useful advice on the numerical simulations. We also thank Katy Clough and Lam Hui for their comments.
V.C.\ is a Villum Investigator and a DNRF Chair, supported by VILLUM FONDEN (grant no.~37766) and by the Danish Research Foundation. V.C.\ acknowledges financial support provided under the European
Union's H2020 ERC Advanced Grant ``Black holes: gravitational engines of discovery'' grant agreement
no.\ Gravitas–101052587.
T.I.\ acknowledges financial support provided under the European Union's H2020 ERC, Starting
Grant agreement no.~DarkGRA--757480.
R.V. was supported by "la Caixa" Foundation grant no. LCF/BQ/PI20/11760032 and Agencia Estatal de Investigación del Ministerio de Ciencia e Innovación grant no. PID2020-115845GB-I00. R.V. also acknowledges support by grant no. CERN/FIS-PAR/0023/2019. 
M.Z.\ acknowledges financial support provided by FCT/Portugal through the IF programme, grant IF/00729/2015, and
by the Center for Research and Development in Mathematics and Applications (CIDMA) through the Portuguese Foundation for Science and Technology (FCT -- Funda\c{c}\~ao para a Ci\^encia e a Tecnologia), references UIDB/04106/2020, UIDP/04106/2020 and the projects PTDC/FIS-AST/3041/2020 and CERN/FIS-PAR/0024/2021.
This project has received funding from the European Union's Horizon 2020 research and innovation programme under the Marie Sklodowska-Curie grant agreement No 101007855.
We thank FCT for financial support through Project~No.~UIDB/00099/2020.
We acknowledge financial support provided by FCT/Portugal through grants PTDC/MAT-APL/30043/2017 and PTDC/FIS-AST/7002/2020.
Computations were performed on the ``Baltasar Sete-Sois'' cluster at IST and XC40 at YITP in Kyoto University
The authors gratefully acknowledge the HPC RIVR consortium and EuroHPC JU for funding this research by providing computing resources of the HPC system Vega at the Institute of Information Science."

\bibliography{References}
\appendix
\section{Supplemental material}
\subsection{Numerical formulation}

There are various numerical formulations of Einstein's equations in spherically symmetric spacetimes.
In spherical symmetry, in addition to the hyperbolicity of the evolution equations, we need to pay attention to numerical instabilities related to the regularity at the origin of the coordinate system~\cite{Alcubierre:2004gn,Arbona:1998hu}.
Herein we use the generalized Baumgarte-Shapiro-Shibata-Nakamura (GBSSN) formulation for the time evolutions~\cite{Shibata:1995we,Baumgarte:1998te,Gundlach:2006tw,Cardoso:2014uka} -- see Ref.~\cite{Brown:2007nt} for the analysis of the hyperbolicity of the system.
Time integration is performed using a 2nd order partially implicit Runge-Kutta method (cf.~\cite{Montero:2012yr}),
and spatial derivatives evaluated using a 4th-order-accurate finite difference method.
We have also implemented the evolution code using the standard 4th order Runge-Kutta method and checked that the results are consistent.
Our numerical code is parallelized with OpenMP and MPI. 

In the GBSSN formulation the 3-metric $\gamma_{ij}$ and the extrinsic curvature $K_{ij}$ are decomposed as
\begin{gather}
\gamma_{ij}=\Phi^{4}\tilde{\gamma}_{ij}\,,\\
K_{ij}=\Phi^{4}\tilde{A}_{ij}+\frac{1}{3}\gamma_{ij}K\,,
\end{gather}
where $\Phi$ is the conformal factor, $\tilde{\gamma}_{ij}$ is the conformal metric,
$K$ is the trace of the extrinsic curvature, and $\tilde{A}_{ij}$ is the traceless part of the extrinsic curvature ($\tilde{\gamma}^{ij}\tilde{A}_{ij}=0$).
In order to uniquely define the conformal factor, we introduce the reference (flat) 3-metric $\bar{\gamma}_{ij}$.
Using the reference metric, we fix the determinant of the conformal 3-metric as follows
\begin{align}
{\rm det}\tilde{\gamma}={\rm det}\bar{\gamma}\,.
\end{align}
%
We also define $\tilde{\Lambda}^{k}$ as
\begin{align}
\tilde{\Lambda}^{k}=\tilde{\gamma}^{ij}(\tilde{\Gamma}^{k}_{ij}-\bar{\Gamma}^{k}_{ij})\,,
\end{align}
where $\tilde{\Gamma}^{k}_{ij}$ and $\bar{\Gamma}^{k}_{ij}$ are the Christoffel symbols for the conformal metric $\tilde{\gamma}_{ij}$
and the reference metric $\bar{\gamma}_{ij}$, respectively.

For spherically symmetric spacetimes we can assume
\begin{gather*}
\bar{\gamma}_{ij}
=\mbox{diag}(\bar{a},\bar{b}r^{2},\bar{b}r^{2}\sin^{2}\theta)\,,\\ \tilde{\gamma}_{ij}
=\mbox{diag}(a,br^{2},br^{2}\sin^{2}\theta)\, ,\\
\tilde{A}^{i}_{j}
=\mbox{diag}(A,B,B)\,,\quad \tilde{\Lambda}^{k}=(\tilde{\Lambda},0,0)\,.
\end{gather*}
From the traceless condition for $\tilde{A}^{i}_{j}$, we can fix $B=-\frac{A}{2}$.
For the evolution of the conformal factor there are two natural choices, the so-called Lagrangian and the Eulerian options.
In order to consider both cases in a unified way we introduce the parameter $\sigma$, where
the Lagrangian case corresponds to $\sigma=1$ and the Eulerian one to $\sigma=0$.
In practice, we evolve the variable $X=\Phi^{-2}$ instead of  $\Phi$.

The evolution equations for each metric variables are
\begin{subequations}
\begin{eqnarray}
\dot{X}&=&\beta X'+\frac{1}{3}\alpha KX-\sigma \frac{1}{3}\mathcal{B}X\,,\\
\dot{a}&=&\beta a^{\prime}+2a\beta^{\prime}-2\alpha aA -\sigma\frac{2}{3}a\mathcal{B}\,,\\
\dot{b}&=&\beta b^{\prime}+2\beta\frac{b}{r}-2\alpha bB -\sigma\frac{2}{3}b\mathcal{B}\,,\\
\dot{A}&=-&\left(\nabla^{r}\nabla_{r}\alpha-\frac{1}{3}\Delta\alpha\right)+\alpha\left(R^{r}_{r}-\frac{1}{3}R\right) \nonumber\\
&&+\beta\partial_{r}A+\alpha K A-\frac{16}{3}\pi\alpha (S^{r}_{r}-S^{\theta}_{\theta})\,,\\
\dot{K}&=&-\Delta\alpha+\beta\partial_{r}K+\alpha\left(A^{r 2}_{r}+2A^{\theta 2}_{\theta}+\frac{1}{3}K^{2}\right) \notag\\
&&+4\pi \alpha(E+S^{r}_{r}+2S^{\theta}_{\theta})\,,
\end{eqnarray}
\end{subequations}
and
\beq
\dot{\tilde{\Lambda}}&=&
\frac{\beta''}{a}
+\frac{\beta'}{\bar{a}}\left(\frac{\bar{b}'}{b}+\frac{\bar{a}'}{2a}+\frac{2\bar{b}}{rb}\right)
\nonumber\\
&&+\beta\left(
-\frac{\bar{a}'^{2}}{2a\bar{a}^{2}}+\frac{\bar{a}'\bar{b}'}{\bar{a}^{2}b}+\frac{\bar{a}''}{2a\bar{a}}-\frac{\bar{b}''}{\bar{a}b}\right. \notag\\
&&\left.~~~+\frac{1}{r}\left(\frac{2\bar{b}\bar{a}'}{\bar{a}^{2}b}-\frac{4\bar{b}'}{\bar{a}b}\right)
-\frac{1}{r^{2}}\frac{2\bar{b}}{\bar{a}b}\right)
\nonumber \notag\\
&&+\frac{\alpha}{a}\left(
\frac{a'}{a}-\frac{\bar{a}'}{\bar{a}}+\frac{b'}{b}-\frac{a\bar{b}'}{\bar{a}b}-\frac{6X'}{X}-\frac{2\alpha'}{\alpha}\right. \notag\\
&&\left.~~+\frac{2}{r}\left(1-\frac{a\bar{b}}{\bar{a}b}\right)
\right)A-\frac{4\alpha K'}{3a}+\frac{\sigma}{3a}\mathcal{B}' \notag\\
&&+\sigma\left(\frac{2}{3}+\chi\right)\left(\frac{a'}{2a^{2}}-\frac{1}{a}\left(\frac{b'}{b}+\frac{2}{r}\right)-\frac{\bar{a}'}{2a\bar{a}}\right. \notag\\
&&\left.~~+\frac{\bar{b}}{\bar{a}b}\left(\frac{\bar{b}'}{\bar{b}}+\frac{2}{r}\right)\right)\mathcal{B}-\sigma\left(\frac{2}{3} +\chi\right)\mathcal{B}\tilde{\Lambda} \notag \\
&&+\beta\tilde{\Lambda}^{\prime}-\tilde{\Lambda}\beta^{\prime}-16\pi\alpha\frac{p_{r}}{a}+\frac{2\sigma}{3}\tilde{\Lambda}\mathcal{B} \notag\,,
\eeq
where
\begin{align*}
\mathcal{B}&=\beta^{\prime}+\left(\frac{a^{\prime}}{2a}+\frac{b^{\prime}}{b}+\frac{2}{r}\right)\beta\,,
\end{align*}
and $R_{ij}$ is the Ricci tensor of the 3-metric $\gamma_{ij}$, whose traceless part can be written as
\begin{align*}
  R^{r}_{r}-\frac{1}{3}R=R^{\rm TF (1)}+X^{2}\bigg(& R^{\rm TF (2)}+R^{\rm TF(3)}\\
 & + \frac{R^{\rm TF(4)}}{r}+\frac{R^{\rm TF(5)}}{r^{2}}\bigg)\,,
\end{align*}
where
\begin{eqnarray*}
R^{\rm TF (1)}&=&\frac{2}{3a}XX''-\frac{2}{3a}\left(\frac{1}{r}+\frac{a'}{2a}+\frac{b'}{2b}\right)XX'\,,\\
R^{\rm TF (2)}&=&\frac{2}{3}\Lambda'+\frac{2}{3}\left(-\frac{1}{r}+\frac{a'}{2a}-\frac{b'}{2b}\right)\Lambda\,,\\
R^{\rm TF (3)}&=&\frac{a'^{2}}{2a^{3}}
-\frac{a'\bar{a}'}{6a^{2}\bar{a}}
-\frac{\bar{a}'^{2}}{3a\bar{a}^{2}}
-\frac{\bar{a}'b'}{6a\bar{a}b}
-\frac{2b'^{2}}{3ab^{2}}
-\frac{a'\bar{b}'}{3a\bar{a}b}
\\
&&+\frac{\bar{a}'\bar{b}'}{3a\bar{a}\bar{b}}
+\frac{b'\bar{b}'}{\bar{a}b^{2}}
+\frac{\bar{b}'^{2}}{3a\bar{b}^{2}}
-\frac{\bar{b}'^{2}}{\bar{a}b\bar{b}}
-\frac{a''}{3a^{2}}
+\frac{\bar{a}''}{3a\bar{a}}\\
&&+\frac{b''}{3ab}
+\frac{2\bar{b}''}{3\bar{a}b}
-\frac{2\bar{b}''}{3a\bar{b}}\,,
\\
R^{\rm TF (4)}&=&-\frac{2\bar{b}a'}{3a\bar{a}b}
+\frac{\bar{a}'}{3a\bar{a}}
-\frac{4b'}{3ab}
+\frac{2\bar{b}b'}{\bar{a}b^{2}}
+\frac{2\bar{b}'}{3\bar{a}b}
-\frac{4\bar{b}'}{3a\bar{b}}\,,\\
R^{\rm TF (5)}&=&-\frac{2}{a}+\frac{2}{3b}+\frac{4\bar{b}}{3\bar{a}b}\,.
\end{eqnarray*}
$\chi$ is a free parameter used to stabilize the simulation~\cite{Yo:2002bm}, which we set to $\chi=-\frac{2}{3}$.
$E$ and $S_{ij}$ are the energy density and stress tensor of the matter sector.

The energy, momentum density, and stress tensor of the complex scalar field are given by
\begin{eqnarray*}
E&=&|\Pi|^{2}+\frac{X^{2}}{a}|\Phi'|^{2}+\mu^{2}|\Phi|^{2}\,,\\
p_{i}&=&-2\left(\partial_{i}\Phi_{\rm R}\Pi_{\rm R}+\partial_{i}\Phi_{\rm I}\Pi_{\rm I}\right)\,,\\
S^{r}_{r}&=&|\Pi|^{2}+\frac{X^{2}}{a}|\Phi'|^{2}-\mu^{2}|\Phi|^{2}\,,\\
S^{\theta}_{\theta}&=&|\Pi|^{2}-\frac{X^{2}}{a}|\Phi'|^{2}-\mu^{2}|\Phi|^{2}\,.
\end{eqnarray*}

We also obtain the Hamiltonian and momentum constraints, as well as the constraint which appears from the definition of $\tilde{\Lambda}^{k}$,
\begin{subequations}
\begin{eqnarray}
H&=&-\frac{X}{2a}\left(
X''+\left(\frac{2}{r}-\frac{a'}{2a}+\frac{b'}{b}\right)X'-\frac{3}{2}\frac{X'^{2}}{X}\right) \notag\\
&&-\frac{X^{2}}{8}\tilde{R}+\frac{A^{2}+2B^{2}}{8}-\frac{K^{2}}{12}+2\pi E\,,\\
M&=&A'-3\frac{X'}{X}A+\frac{3}{r}A+\frac{3b'}{2b}A-\frac{2}{3}K'-8\pi p\,,\\
\tilde{\Lambda}&=&\frac{a'}{2a^{2}}-\frac{1}{a}\left(\frac{b'}{b}+\frac{2}{r}\right)-\frac{\bar{a}'}{2a\bar{a}}+\frac{\bar{b}}{\bar{a}b}\left(\frac{\bar{b}'}{\bar{b}}+\frac{2}{r}\right)\,,\nn\\
\end{eqnarray}
\end{subequations}
where $\tilde{R}$ is the Ricci scalar with respect to $\tilde{\gamma}$, given by
\[
  \tilde{R}=R^{(1)}+\frac{R^{(2)}}{r^{2}}+\frac{R^{(3)}}{r}+R^{(4)}\,,
\]
with
\begin{align*}
R^{(1)}&=\Lambda'+\left(\frac{2}{r}+\frac{b'}{b}+\frac{a'}{2a}\right)\Lambda\,,\quad R^{(2)}=-\frac{2}{b}\left(1-\frac{\bar{b}}{\bar{a}}\right)\,,\\
R^{(3)}&=-\frac{\bar{b}a'}{a\bar{a}b}-\frac{\bar{a}'}{a\bar{a}}-\frac{2b'}{ab}+\frac{4\bar{b}'}{\bar{a}b}+\frac{4\bar{b}'}{a\bar{b}}\,,\\
R^{(4)}&=\frac{3a'^{2}}{4a^{3}}-\frac{a'\bar{a}'}{4a^{2}\bar{a}}-\frac{\bar{a}'^{2}}{2a\bar{a}^{2}}+\frac{\bar{a}'b'}{2a\bar{a}b}+\frac{b'^{2}}{2ab^{2}}\,,\\
&-\frac{a'\bar{b}'}{2a\bar{a}b}-\frac{\bar{a}'\bar{b}'}{a\bar{a}\bar{b}}-\frac{\bar{b}'^{2}}{a\bar{b}^{2}}-\frac{a''}{2a^{2}}+\frac{\bar{a}''}{2a\bar{a}}-\frac{b''}{ab}+\frac{\bar{b}''}{\bar{a}b}+\frac{2\bar{b}''}{a\bar{b}}.
\end{align*}

Finally, from the Klein-Gordon equation we obtain the evolution equations for the complex scalar field
\begin{subequations}
\begin{eqnarray}
\dot{\Phi}&=&\alpha \Pi+\beta\Phi'\,,\\
\dot{\Pi}&=&\alpha \Delta\Phi+D\alpha D\Phi-\mu^{2}\alpha\Phi+\beta\partial_{r}\Pi+\alpha K\Pi .
\end{eqnarray}
\end{subequations}

We use the standard ``1+log'' and ``hyperbolic gamma driver'' gauge condition
\be
\dot{\alpha}=\beta\partial_{r}\alpha-2K\alpha\,,\quad \dot{\beta}=k_{1}\lambda\,,
\ee
where $\lambda$ is an auxiliary field with evolution equation given by
\begin{align}
\dot{\lambda}&=k_{2}\Lambda \,.
\end{align}
$k_{1}$ and $k_{2}$ are constants which we set to $k_{1}=1$ and $k_{2}=\frac{3}{4}$.
During the evolution, we monitor the apparent horizon area radius $r_{\rm AH}$ to calculate the BH irreducible mass $M=\frac{r_{\rm AH}}{2}$.

\subsection{Construction of initial data}

We now summarize the construction of the BS-BH initial data. We assume momentarily static and conformally flat initial data, 
with vanishing shift vector and  vanishing  time derivatives of the shift and lapse functions:
\beq
a(0,r)&=&\bar{a}(r)\,,\qquad b(0,r)=\bar{b}(r)\,,\\
A(0,r)&=&B(0,r)=K(0,r)\nonumber\\
&=&\beta(0,r)=\dot{\beta}(0,r)=\dot{\alpha}(0,r)=0\,.
\eeq
where $\bar{a}$ and $\bar{b}$ are the reference metric values. The conformally flat condition implies
\begin{align}
  \bar{a}=r_{\ast}'(r)^{2}\,,\qquad
  \bar{b}=\left(\frac{r_{\ast}(r)}{r}\right)^{2} \,,
\end{align}
where the function $r_{\ast}(r)$ is the coordinate transformation from $r$ to $r_{\ast}$, in which the initial conformal line element is $dr_{\ast}^{2}+r_{\ast}^{2}d^{2}\Omega$.
Under these assumptions, the momentum constraint is trivially satisfied and we merely need to solve the Hamiltonian constraint.

To solve the Hamiltonian constraint equation we first need to specify our free data, which in our case corresponds to a superposition of the configurations of a BS with that of a BH, in isotropic coordinates. For the BS data, we assume a harmonic ansatz for the scalar field,
\[
\psi_{\rm BS}=\psi_{0,\rm BS}(r)e^{-i\omega t}\,,
\]
where $\psi_{0}(r)$ is a real scalar function and~$\omega$ is the BS frequency, which is determined from the boundary conditions for the given $\psi_{0,\rm BS}(0)$.
We write the metric ansatz as
\begin{equation}
ds^{2}=-\alpha_{\rm BS}^{2}dt^{2}+\Phi_{\rm BS}^{4}\left(
\bar{a}(r)dr^{2}+\bar{b}(r)r^{2}d^{2}\Omega
\right)\,.
\label{metric_isotropic_2}
\end{equation}
The corresponding equations are
\begin{subequations}
\begin{align}
\psi_{0,\rm BS}''&=\left(
-\frac{2}{r}
+\frac{\bar{a}'}{2\bar{a}}
-\frac{\bar{b}'}{\bar{b}}
-2\frac{\Phi_{\rm BS}'}{\Phi_{\rm BS}}
-\frac{\alpha_{\rm BS}'}{\alpha_{\rm BS}}
\right)\psi_{0,{\rm BS}}'
\nonumber\\
&+\bar{a}\left(
-\frac{\omega^{2}}{\alpha_{\rm BS}^{2}}+\mu^{2}
\right)\Phi^{4}\psi_{0,{\rm BS}}\,,
\label{eq:ODE psi0} \\
\Phi_{\rm BS}''&=\left(
-\frac{2}{r}+\frac{\bar{a}'}{2\bar{a}}-\frac{\bar{b}'}{\bar{b}}
\right)\Phi_{\rm BS}'-2\pi\Phi_{\rm BS}^{5}\bar{a}E_{\rm BS}\,,
\label{Ham eq BS} \\
\frac{\alpha_{\rm BS}'}{\alpha_{\rm BS}}
&=
\bigg(r\Phi_{\rm BS} \bar{b}'+2\bar{b}(\Phi_{\rm BS}+2r\Phi_{\rm BS}')\bigg)^{-1}\times
\nonumber\\
&\left\{
\left(
\frac{\bar{a}-\bar{b}}{r}-\bar{b}'-\frac{r\bar{b}'^{2}}{4\bar{b}}\right)
-(4\bar{b}+2r\bar{b}')\Phi_{\rm BS}'
-4r\bar{b}\frac{\Phi_{\rm BS}'^{2}}{\Phi_{\rm BS}}
\right.
\nonumber\\
&\left.
+8\pi r\bar{b}\left(
-\bar{a}\left(\mu^{2}-\frac{\omega^{2}}{\alpha_{\rm BS}^{2}}
\right)\psi_{0,{\rm BS}}^{2}\Phi_{\rm BS}^{5}
+\Phi_{\rm BS}\psi_{0,\rm{BS}}'^{2}\right)
\right\}.
\nonumber\\
\label{eq:ODE lap}
\end{align}
\label{eq:ODE BS}
\end{subequations}
The energy density is given by
\begin{equation}
  E_{\rm BS}=\frac{\psi_{0,{\rm BS}}'^{2}}{\bar{a}\Phi_{\rm BS}^{4}}+\left(\mu^{2}+\frac{\omega^{2}}{\alpha_{\rm BS}^{2}}\right)\psi_{0,{\rm BS}}^{2}\,.
  \label{eq:EBS}
\end{equation}

The construction of a BS is an eigenvalue problem for the BS frequency $\omega$, imposing regularity at the origin and vanishing scalar field at spatial infinity.
When integrating Eq.~\eqref{eq:ODE psi0} for $\psi_{0,\rm BS}$ from the origin, 
the exponentially growing mode dominates, making it 
difficult to construct the solution for large coordinate radii.
One possible way to construct a BS solution for a large numerical domain consists in integrating Eq.~\eqref{eq:ODE psi0} until a certain radius
and extrapolating the solution by hand until the numerical boundary.
This is simple to do, but one can not avoid artificially exciting some modes.
Here we present a better alternative, using a combination between a shooting and a relaxation method.

After constructing the solution and obtaining the BS frequency by a shooting method until a certain radius with a fixed $\psi_{0,\rm BS}(0)$,
we solve Eq.~\eqref{eq:ODE psi0} again, using a successive over-relaxation (SOR) method for Eqs.~\eqref{Ham eq BS} and \eqref{eq:ODE lap} in the full numerical domain,
to obtain $\Psi_{\rm BS}$, $\alpha_{\rm BS}$, and $\psi_{0,\rm BS}$.
During the relaxation procedure, we fix the frequency $\omega$ and find $\psi_{0,\rm BS}(0)$ such that the two boundary conditions are consistent with the given frequency.
After the relaxation we check that the new value of $\psi_{0,\rm BS}(0)$ is not too different from its original value (typically, there is less than 1\% difference).

Now we want to solve the Hamiltonian constraint for a configuration with this scalar field profile and a BH at its center. The equation for the 3-metric conformal factor is
\begin{equation}
\Phi''+\left(
\frac{2}{r}-\frac{\bar{a}'}{2\bar{a}}+\frac{\bar{b}'}{\bar{b}}
\right)\Phi'+2\pi\Phi^{5}\bar{a}E=0\,,
\end{equation}
where
\[
E=\frac{\psi_{0,{\rm BS}}'^{2}}{a\Phi^{4}}+\left(\mu^{2}+\frac{\omega^{2}}{\alpha_{\rm BS}^{2}}\right)\psi_{0,{\rm BS}}^{2} \,.
\]
To construct the initial configuration with a BH, we superpose the conformal factor
of the BH spacetime with mass $M_{0}$
and the conformal factor of the BS solution in isotropic coordinates.
This superposition is not a solution of the constraint equation, so we introduce a correction term $\delta\Phi$ to the conformal factor,
\begin{align}
\Phi=\Phi_{\rm BS}+\Phi_{\rm BH}+\delta\Phi\,,\label{def deltaPhi}
\end{align}
where $\Phi_{\rm BH}=\frac{M_{0}}{2r}$, and solve the corresponding elliptic equation for $\delta \Phi$ with boundary conditions $\delta\Phi(\infty)=\delta\Phi'(\infty)=0$.
The equation for $\delta\Phi$ is
\begin{align}
 \delta\Phi''&+\left(
\frac{2}{r}-\frac{\bar{a}'}{2\bar{a}}+\frac{\bar{b}'}{\bar{b}}
\right)\delta\Phi'
+2\pi \bar{a}\left(\mu^{2}+\frac{\omega^{2}}{\alpha_{\rm BS}^{2}}\right)\psi_{0,{\rm BS}}^{2}\Phi^{5}
\nonumber\\
&+2\pi \psi'^{2}_{0,{\rm BS}}\Phi
-2\pi \bar{a}\Phi_{\rm BS}^{5}E_{\rm BS}=0\,,
\label{eq:deltaPhi}
\end{align}
where $\Phi$ is given by~\eqref{def deltaPhi}, $E_{\rm BS}$ by~\eqref{eq:EBS}, and the solution of Eqs.~\eqref{eq:ODE BS} is used. To derive Eq.~\eqref{eq:deltaPhi}, we used Eq.~\eqref{Ham eq BS} for $\Phi_{\rm BS}$. For (Newtonian) BSs considerably larger and heavier than the BH, the correction term~$\delta \Phi$ is relatively small, allowing us to interpret the initial data as still describing a BS with a central BH.

\subsection{Coordinate transformation}

In our setup we have two typical length scales, the BH horizon radius 
and the Compton wavelength of the scalar field.
For simulations with large BSs, these two length scales are very different
and we must resolve both of them during the numerical evolution.
Therefore, to accurately resolve both scales, we use the coordinate transformation
\begin{align*}
\partial_{r_{\ast}}r(r_{\ast})=
\begin{cases}
1       &(r_{\ast}\leq r_{\ast \rm in})\\
g(r_{\ast};r_{\ast\rm in},r_{\ast\rm out},s)    &(r_{\ast\rm in}< r_{\ast}\leq r_{\ast\rm out})\\
s        &(r_{\ast\rm out}< r_{\ast})
\end{cases}
\end{align*}
%
with~$r_{\ast}$ the ``old''
and~$r$ the ``new'' radial coordinate; $g(r_{\ast};r_{\ast\rm in},r_{\ast\rm out},s)$ is the 9th order smooth polynomial satisfying
\begin{gather*}
g(r_{\ast\rm in};r_{\ast\rm in},r_{\ast\rm out},s)=1,
\\
g(r_{\ast\rm out};r_{\ast\rm in},r_{\ast\rm out},s)=s,
\\
\partial_{r_{\ast}}^{n}g(r_{\ast\rm in};r_{\ast\rm in},r_{\ast\rm out},s)=
\partial_{r_{\ast}}^{n}g(r_{\ast\rm out};r_{\ast\rm in},r_{\ast\rm out},s)=0
\\
(n=1,2,3,4),
\end{gather*}
where~$r_{\ast\rm in}$ and $r_{\ast\rm out}$ are the inner and outer radius of the transformation,
and~$s$ is the ratio between the resolutions in~$r$ and~$r_{\ast}$.
Typically, we set~$r_{\ast\rm in}=100M_{0}$,~$r_{\ast\rm out}=300M_{0}$,
and~$s=20$.
\subsection{Numerical convergence}
\begin{figure*}[thpb]
	\includegraphics[width=0.475\textwidth]{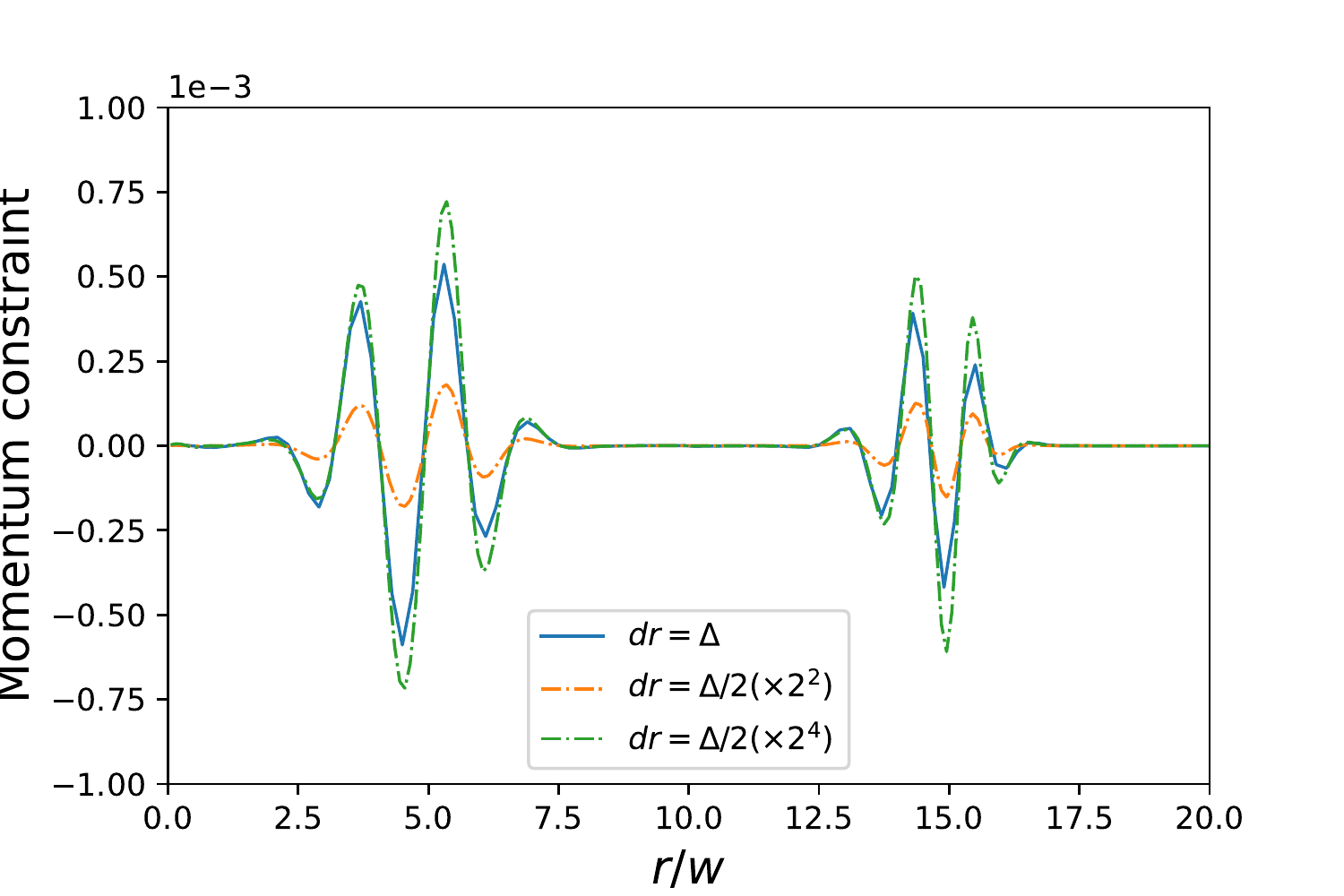} 
	\caption{Momentum constraint at $t=10w$ for pure gauge evolution with initial data given by \eqref{eq:pure gauge ID}.
	\label{Conv. pure gauge}}
\end{figure*}

\begin{figure*}[thpb]
	\includegraphics[width=0.475\textwidth]{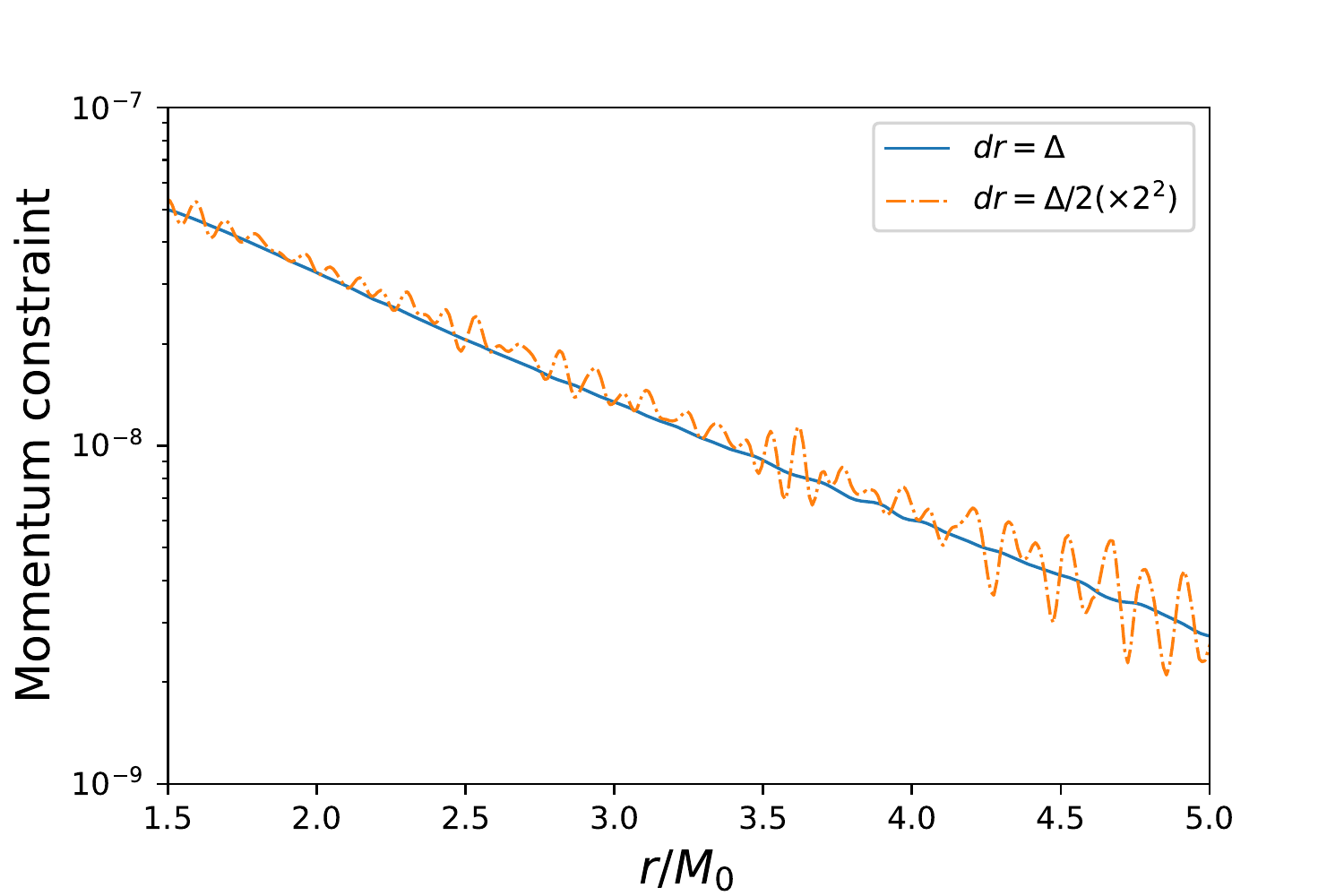} 
		\includegraphics[width=0.475\textwidth]{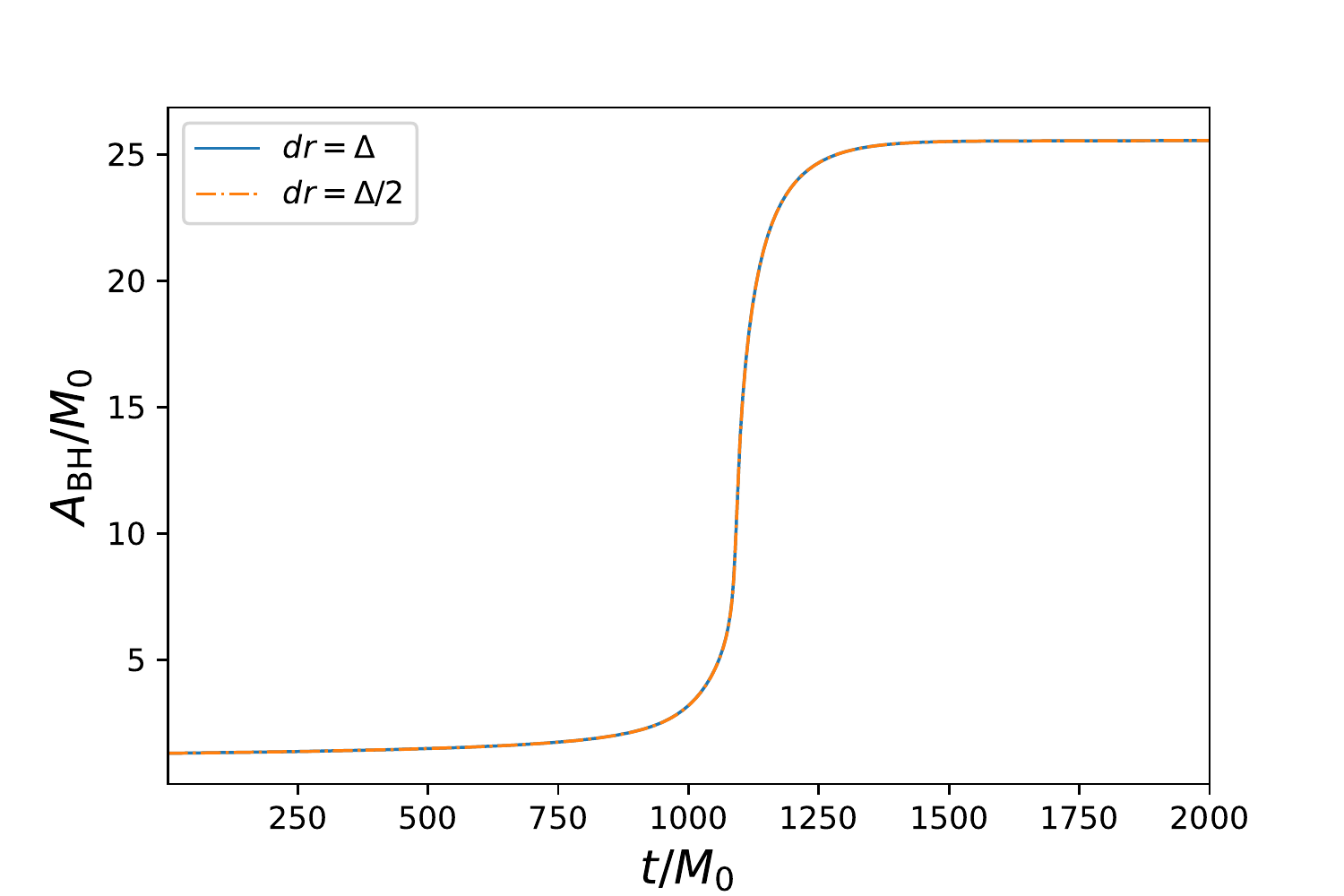} 
	\caption{Left: momentum constraint at $t=800M_{0}$ for a time evolution starting from a BH and BS with $\psi_{0}=0.03$ and $\mu M_{0}= 0.05$. Right: the respective time evolution of the black hole area radius.
	\label{Conv. BSBH}}
\end{figure*}

To check the validity of our numerical code, we test the numerical convergence for two simple setups.
The first setup is corresponds to a gauge evolution on a vacuum flat spacetime \cite{Montero:2012yr}, whose initial data is
\begin{align}
&\alpha=1+\frac{\alpha_{0}r^{2}}{r^{2}+w^{2}}\left(
e^{-\left(\frac{r-r_{0}}{w}
\right)^{2}}+e^{-\left(\frac{r+r_{0}}{w}
\right)^{2}}\right)\,,\\
&\beta^{r}=0\,\\
&a=b=1\,,\\
&A=B=K=0\,,\\
&X=1\,,
\label{eq:pure gauge ID}
\end{align}
where~$\alpha_{0}$, $w$, and $r_{0}$ are, respectively, the amplitude, width, and radius of the initial gauge pulse. 
We fix~$\alpha_{0}=0.01$, $w=1$, and $r_{0}=5$.
Figure \ref{Conv. pure gauge} shows the constraint violation for different resolutions, 
and we confirmed that the time evolution converges with order between 2nd and 4th.
The second test simulation corresponds to the time evolution of the BH and BS with $\psi_{0}=0.03$ and $\mu M_{0}= 0.05$.
Figure \ref{Conv. BSBH} shows the constraint violation outside the BH horizon (left panel) and black hole area radius (right panel) for different resolutions.
From the left panel we can confirm the 2nd order numerical convergence, as expected.
In the right panel we cannot see the difference between the two resolutions in the black hole area radius, concluding that our numerical simulations sufficiently converge.
%

\subsection{Selected simulations}
\begin{figure*}[thpb]
	\includegraphics[width=0.475\textwidth]{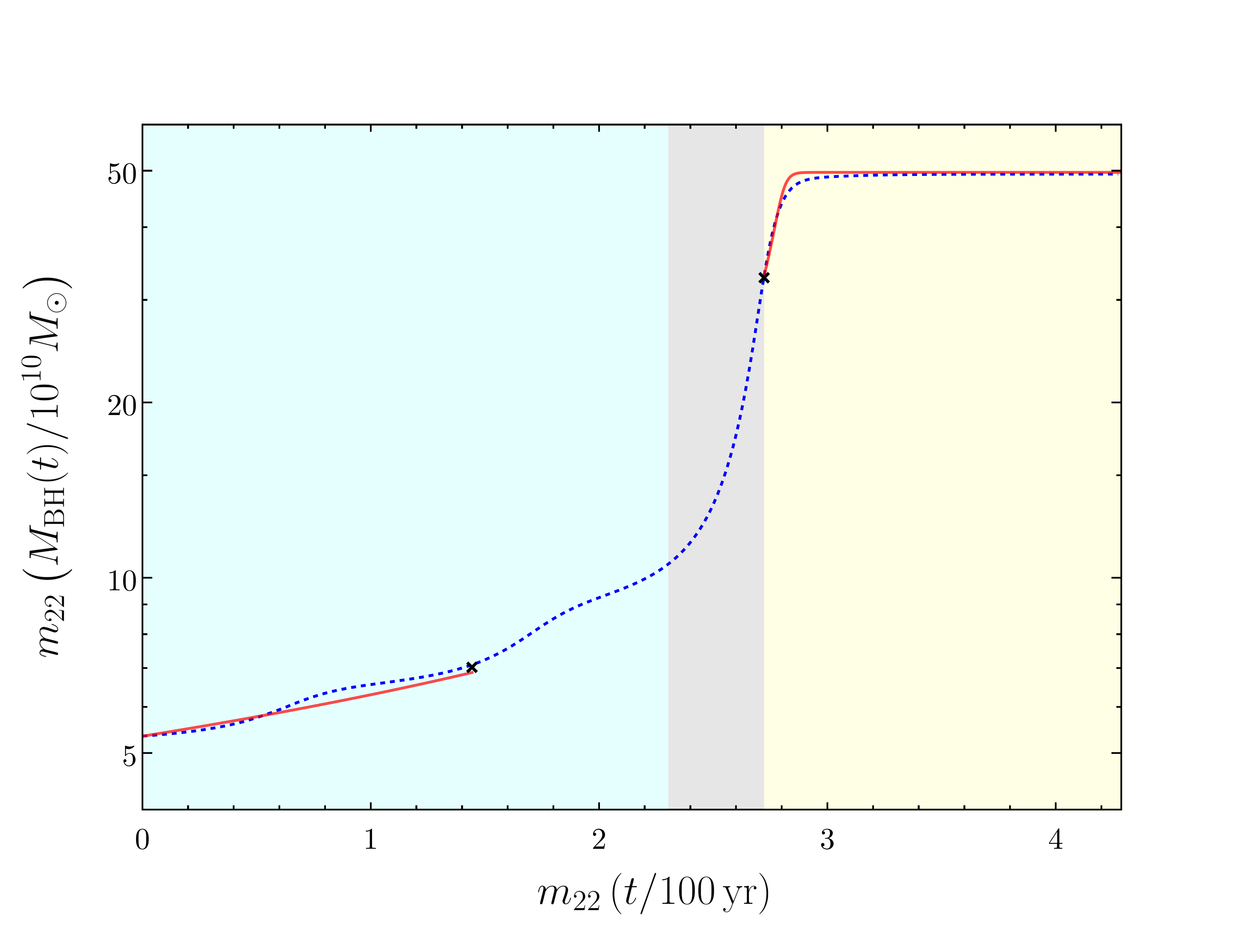} \hfill
	\includegraphics[width=0.475\textwidth]{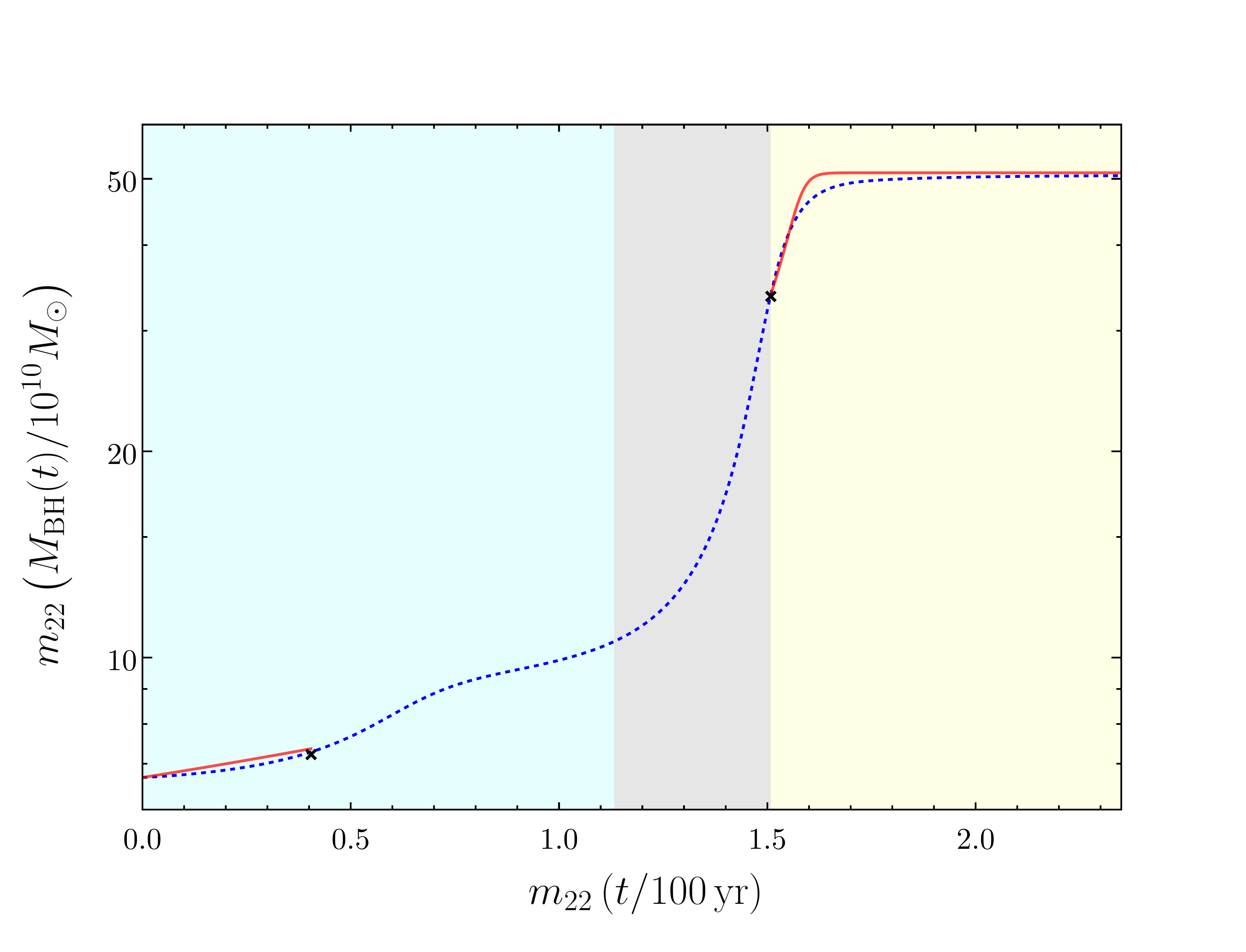}
	\caption{BH mass as function of time (dashed blue) for a BS with initial mass~$M_\text{BS,0}\approx 4\times 10^{11}M_\odot/ m_{22}$ and different initial BH masses (\textbf{left:}~$M_\text{BH,0}\approx5\times 10^{10}M_\odot / m_{22}$; \textbf{right:}~$M_\text{BH,0}\approx7\times 10^{10}M_\odot / m_{22}$). The red curves show the analytical approximations~(11) and~(14), with the black crosses signaling~$\nu=\{\tfrac{1}{6},2\}$ (where, respectively,~(11) ceases and~(14) starts to be valid). We can see the analytics approximate quite well the numerical results, even though the soliton is only marginally Newtonian.}
	\label{accr_psi0005}
\end{figure*}

\begin{figure*}[thpb]
	\includegraphics[width=0.475\textwidth]{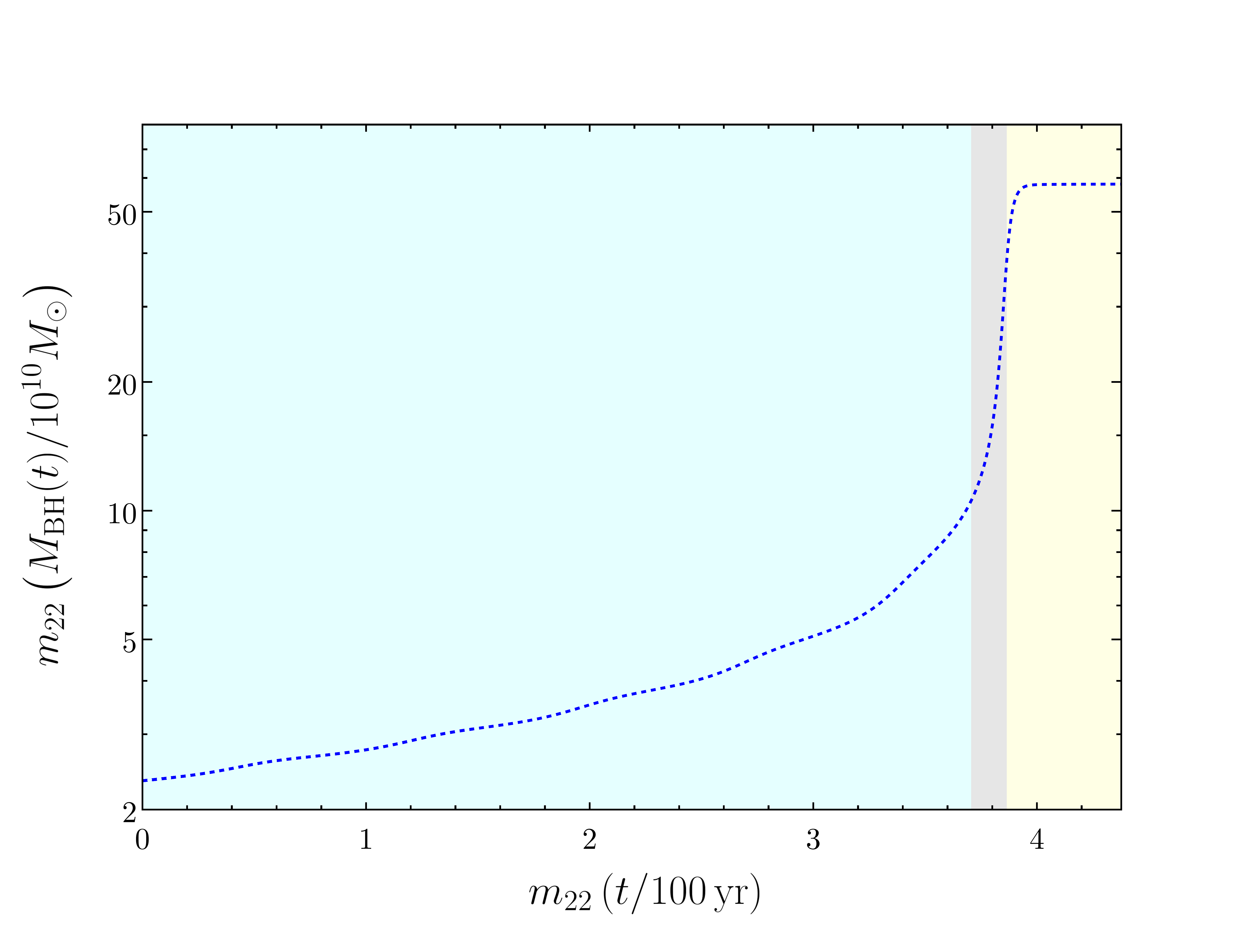} \hfill
	\includegraphics[width=0.475\textwidth]{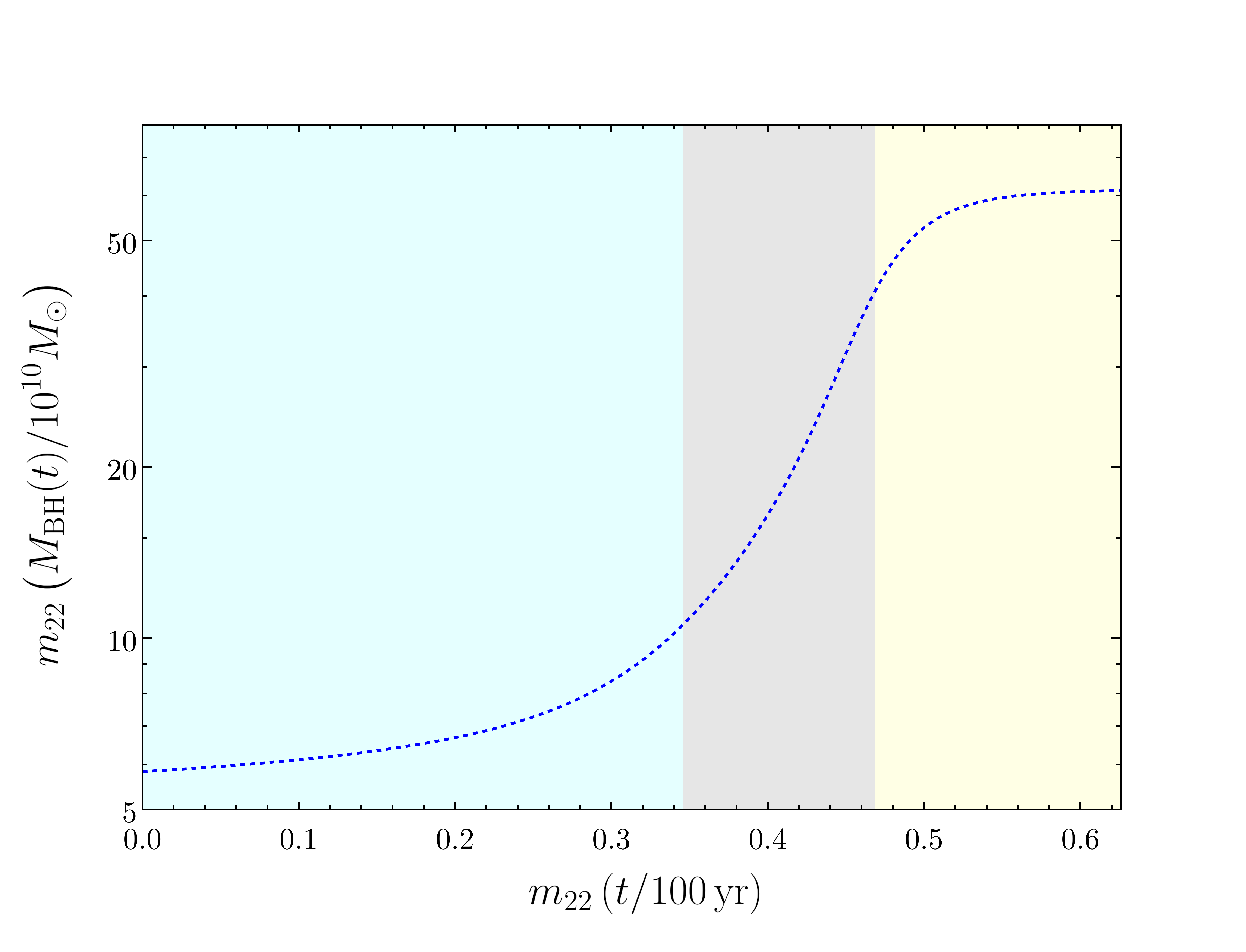}	
	\caption{BH mass as function of time (dashed blue) for a BS with initial mass~$M_\text{BS,0}\approx 6\times 10^{11}M_\odot/ m_{22}$ and different initial BH masses (\textbf{left:}~$M_\text{BH,0}\approx2\times 10^{10}M_\odot / m_{22}$; \textbf{right:}~$M_\text{BH,0}\approx6\times 10^{10}M_\odot / m_{22}$). This soliton is considerably relativistic, so we omit the analytic approximations.}
	\label{accr_psi001}
\end{figure*}

\begin{figure*}[thpb]
		\includegraphics[width=0.475\textwidth]{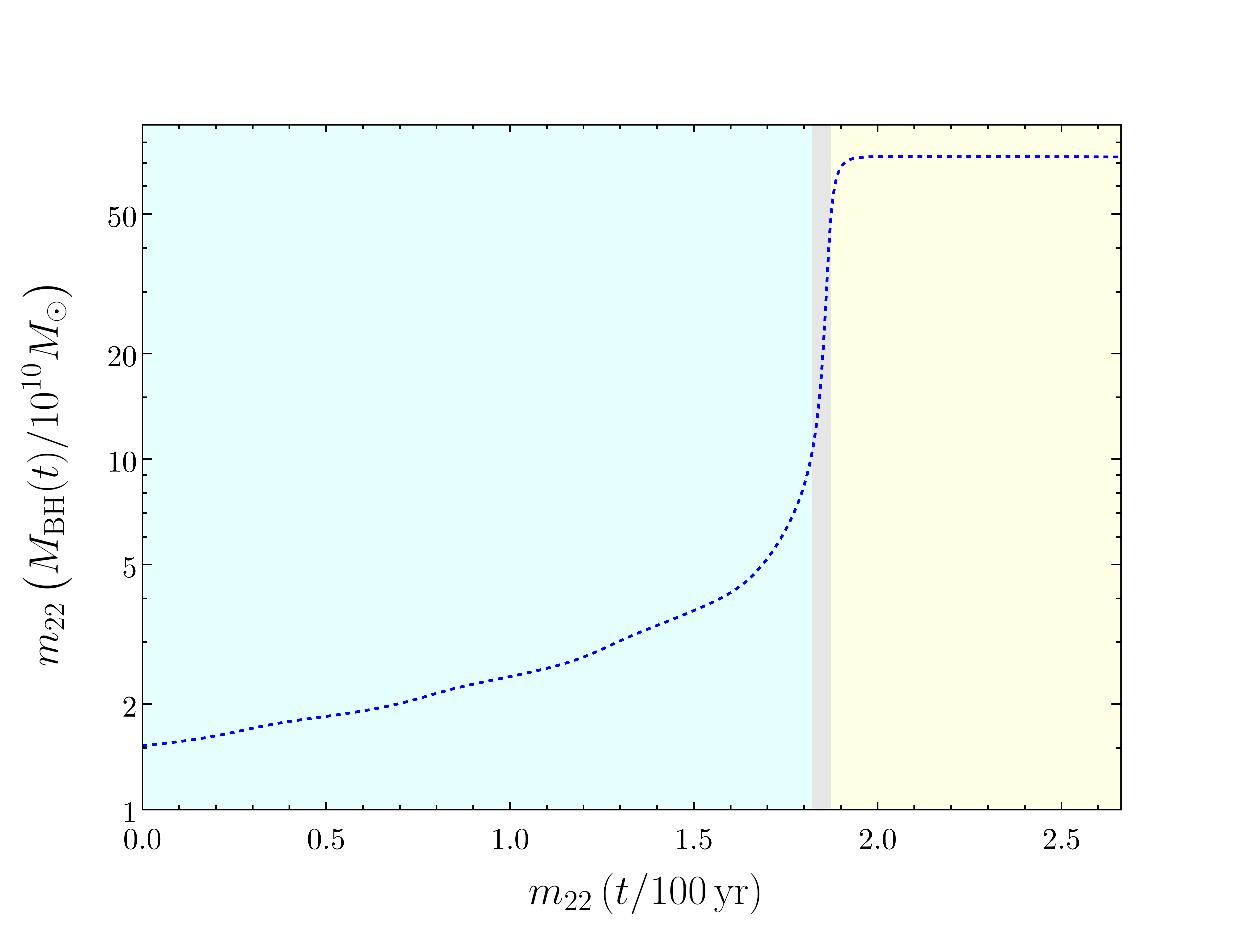} \hfill
		\includegraphics[width=0.475\textwidth]{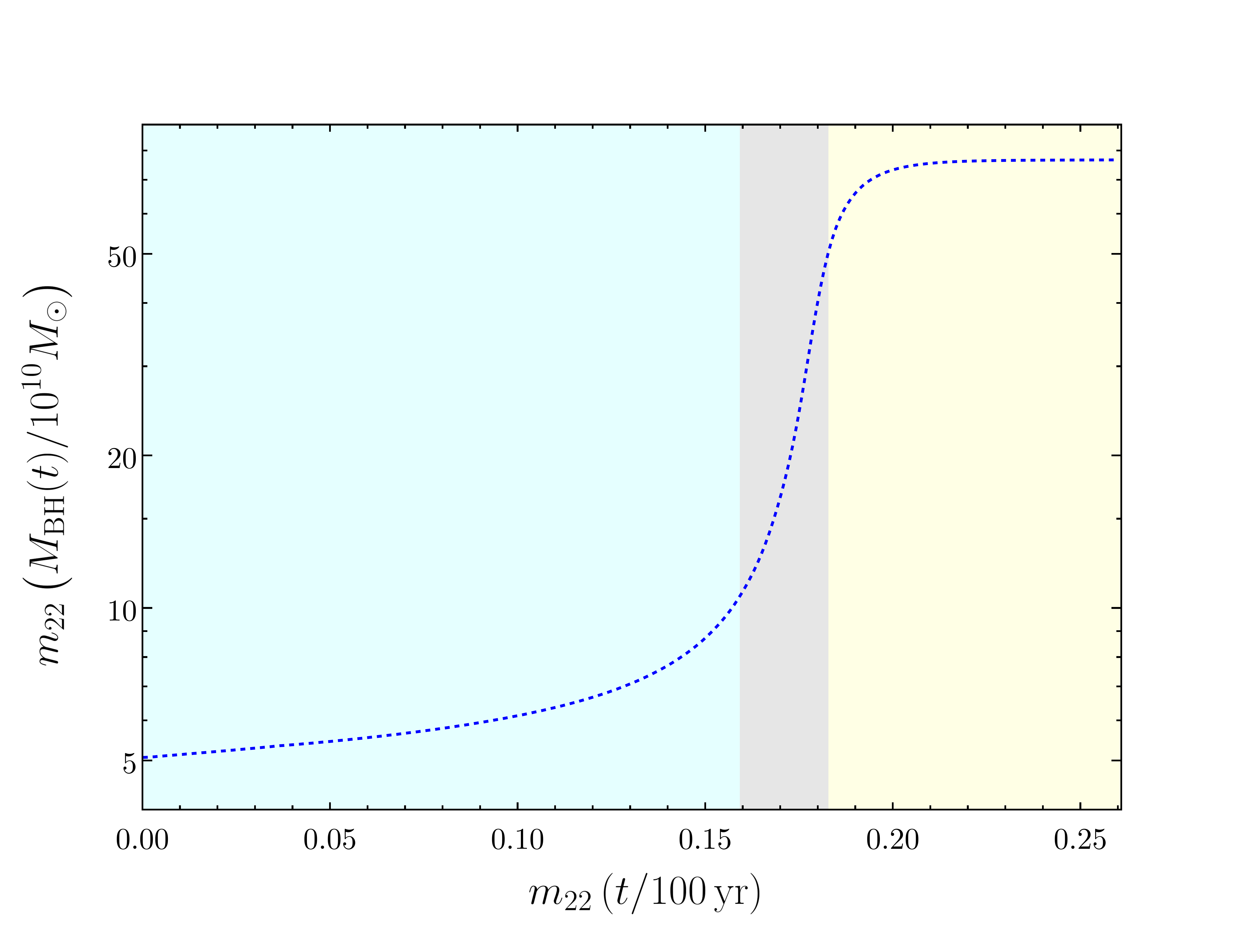}	
	\caption{BH mass as function of time (dashed blue) for a BS with initial mass~$M_\text{BS,0}\approx 6\times 10^{11}M_\odot / m_{22}$ and different initial BH masses (\textbf{left:}~$M_\text{BH,0}\approx3\times 10^{10}M_\odot / m_{22}$; \textbf{right:}~$M_\text{BH,0}\approx7\times 10^{10}M_\odot / m_{22}$). This soliton is relativistic, so we omit the analytic approximations.}
	\label{accr_psi002}
\end{figure*}

\begin{figure*}[thpb]
	\includegraphics[width=0.475\textwidth]{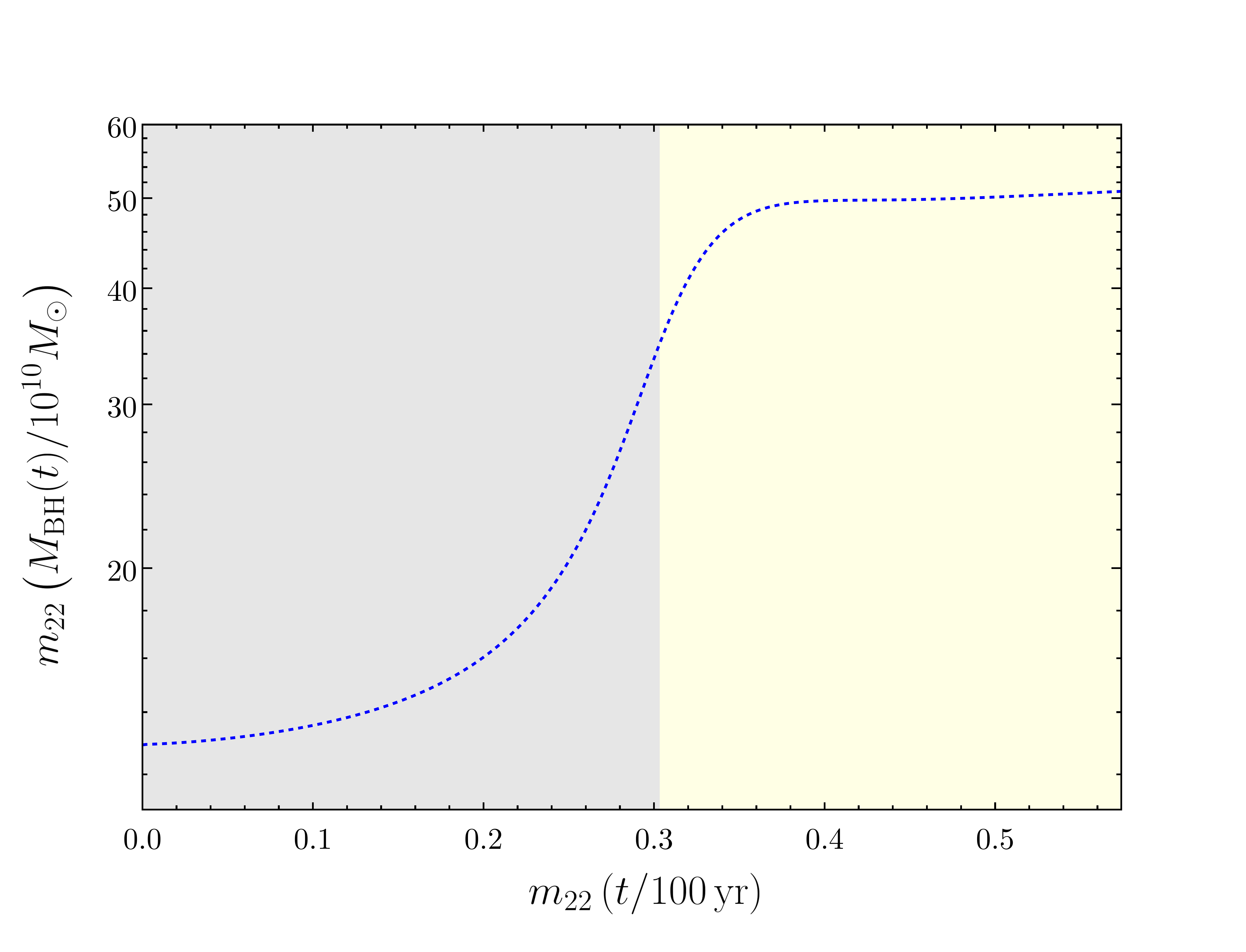} \hfill
	\includegraphics[width=0.475\textwidth]{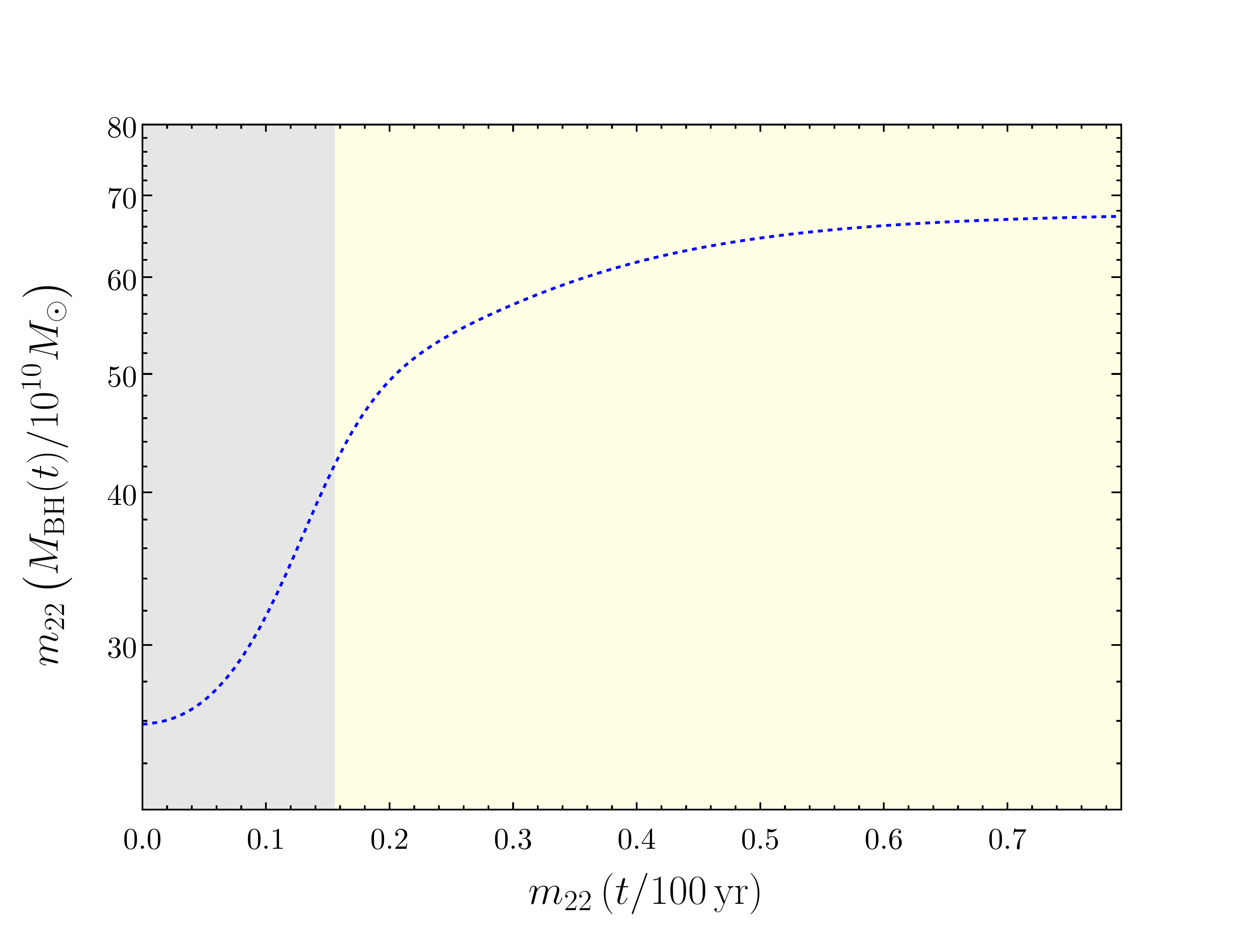}	
	\caption{BH mass as function of time (dashed blue) for a BS with initial mass~$M_\text{BS,0}\approx 4\times 10^{11}M_\odot/ m_{22}$ and different initial BH masses (\textbf{left:}~$M_\text{BH,0}\approx13\times 10^{10}M_\odot / m_{22}$; \textbf{right:}~$M_\text{BH,0}\approx26\times 10^{10}M_\odot / m_{22}$). While this soliton is marginally Newtonian, it hosts a BH of mass comparable to its own. Because of that, this process starts already in stage~II, with the accretion time being of the order of the free-fall time.}
	\label{accr_psi0005_massive}
\end{figure*}

Figures~\ref{accr_psi0005},~\ref{accr_psi001},~\ref{accr_psi002}, and~\ref{accr_psi0005_massive} show the evolution of the BH mass obtained from our numerical relativity simulations for some different initial parameters. In Fig.~\ref{accr_psi0005}, we compare the numerical results with the predictions from the analytical models~(11) and~(14) for the lightests solitons. The results show that, as expected, the accuracy of the analytic results is best in the limit of Newtonian (less massive) solitons and smaller initial mass ratios~$\nu_0$. Figure~\ref{accr_psi0005_massive} shows two simulations with~$\nu_0\sim \mathcal{O}(1)$, where the process starts already at stage~II and the accretion time is dictated by the free-fall time.

\subsection{Projection of a BS onto gravitational atom states}

In order to estimate the support of the scalar field over the different gravitational atom states in the late-time stage of accretion, here we project a ground-state BS onto the gravitational atom states. Since the BS is spherically symmetric, the scalar has support only on~$\ell=0$ states.

As a warm up, note that the size of the gravitational atom states follows~$R_n\sim n^2/(\mu^2M_\text{BH})$ with~$n\geq1$ (where we are using the average radius in the $n$-state), which implies
\be
\frac{R_{\rm BS}}{R_n}\sim \frac{10}{n^2}\frac{M_{\rm BH}}{M_{\rm BS}}\,,
\ee
where~$R_\text{BS}$ is the~$R_{98}$ for the ground-state BS~\cite{Annulli:2020lyc}.
This indicates that for small~$\nu \equiv M_\text{BH}/M_\text{BS}$ the scalar field has support only in the lowest states. 
 
For light scalars $\mu^{2}M_{\rm BH}\ll 1$, the gravitational atom states are well described by~\cite{Brito:2015oca}
\be
\bar{\chi}_n(r)=-2\frac{\left(M_{\rm BH}\mu^2\right)^{3/2}}{n^{5/2}}e^{-\tilde{r}/2}L_{n-1}^{(1)}(\tilde{r})\,,
\ee
where $L_{n-1}^{(1)}(\tilde{r})$ is the Laguerre polynomial~\cite{NIST:DLMF}, and~$\tilde{r}\equiv2M_{\rm BH}\mu^2 r/n$.
These states are spherically symmetric and normalized such that
\be
\int dr\, r^2\bar{\chi}_n \bar{\chi}_{n'}=\delta_{nn'}\,, \notag
\ee
with $\delta_{ij}$ the Kronecker symbol.
In the main text, we used the states~$\chi_n$ normalized such that their mass is equal to~$M_\text{BS}$; these are related to the above states by
\be
\chi_{n}=-\frac{A_{n}n^{3/2}}{2(\mu^{2}M_{\rm BH})^{3/2}}\bar{\chi}_{n}\,,
\ee
with~$A_n$ given in~Eq.(13).

\begin{figure}[thbp]
	\centering
	\includegraphics[width=0.49\textwidth]{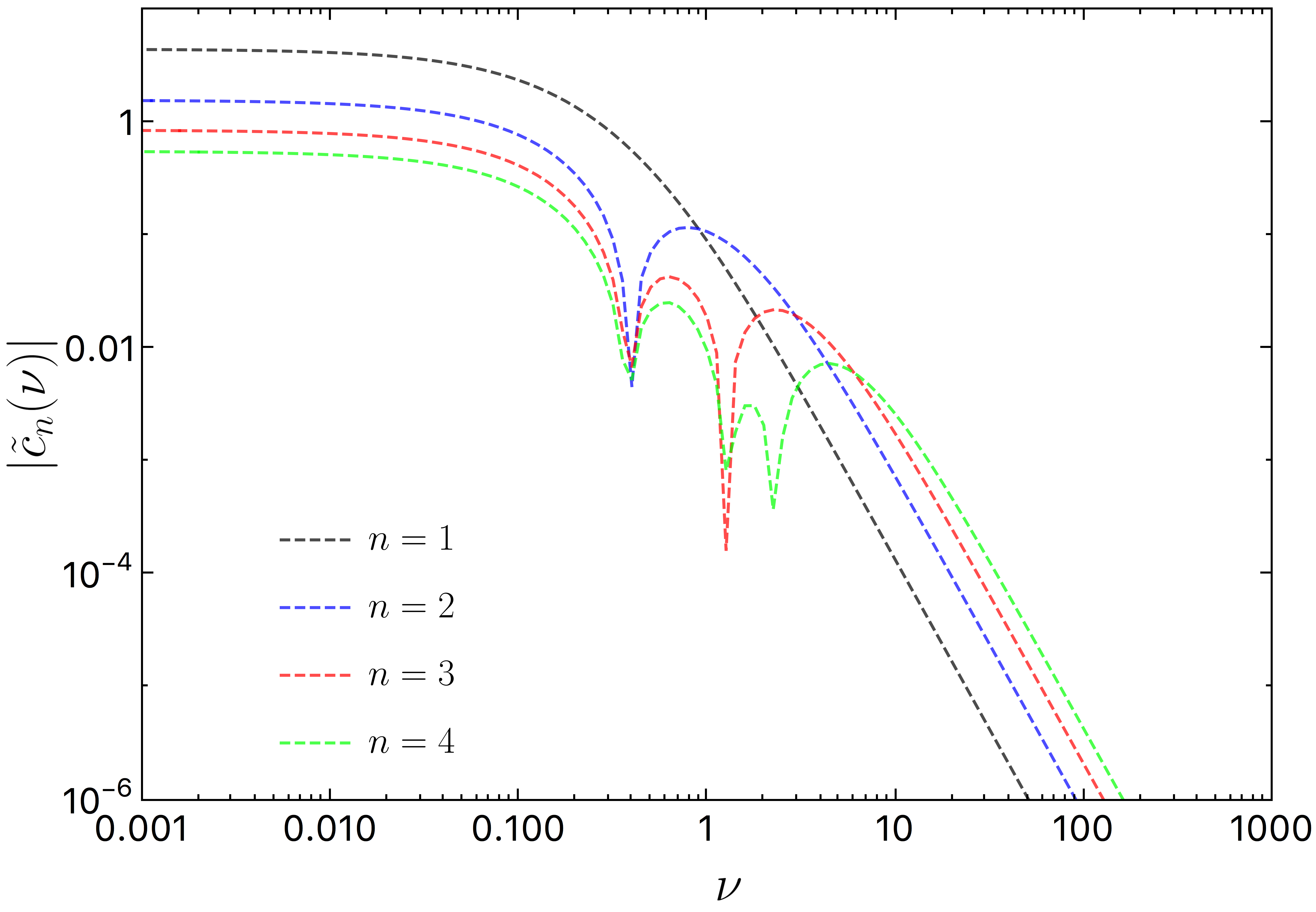}
	\caption{The coefficients~$\tilde{c}_n$ obtained from the projection of a ground-state BS onto the gravitational atom states.
	\label{overlap bs atom}}
\end{figure}

We may then expand the BS as 
\begin{gather}
\chi_{\rm BS}=\sum_{n} \bar{c}_{n}\bar{\chi}_{n}\,,\\
\bar{c}_n=\int dr r^2 \chi_{\rm BS}\, \bar{\chi}_n\,. \notag
\end{gather}
For a Newtonian BS, one has~$\chi_{\rm BS}=\mu^2M_{\rm BS}^2\,g(\mu^2M_{\rm BS} r)$, where the function~$g$ is defined in Eq.~(49) of Ref.~\cite{Annulli:2020lyc}.
Thus, we can write
\begin{gather}
\bar{c}_{n}=-2\frac{\sqrt{M_{\rm BS}}}{\mu}\nu^{3/2}\tilde{c}_{n}(\nu)\,, \nonumber\\
\tilde{c}_{n}(\nu)=\frac{1}{n^{5/2}}\int_{0}^{\infty}dxx^{2}g(x)e^{-\nu x/n}L_{n-1}^{(1)}\left(\frac{2\nu}{n}x\right)\,.
\nonumber
\end{gather}
The first four coefficients~$\tilde{c}_n$ for~$\nu=\{0,0.1,1,10\}$ are presented in the following table.
\begin{center}
\begin{tabular}{c||c|c|c|c}
	$\nu$ & $\tilde{c}_1$ & $\tilde{c}_2/\tilde{c}_1$ & $\tilde{c}_3/\tilde{c}_1$ & $\tilde{c}_4/\tilde{c}_1$ \\ \hline \hline 
	$0$ & $4.4$ & $0.35$ & $0.19$  & $0.12$ \\ 
	$0.1$ & $2.3$ & $0.32$ & $0.18$ & $0.11$ \\
	$1$ & $8.7 \times 10^{-2}$ &  $-1.2$ & $-0.21$ & $-0.11$ \\
	$10$ & $1.3 \times 10^{-4}$ & $-5.4$ & $13$ & $-72$
\end{tabular}
\end{center}
Asymptotically, we find $\tilde{c}_n\propto n^{-3/2}$ for small $\nu$. 
Figure~\ref{overlap bs atom} shows~$\tilde{c}_{n}$ as function of~$\nu$ for $n\leq 4$,
confirming that the fundamental~$n=1$ mode has most of the support for small~$\nu$. 

\end{document}